\documentclass[fleqn,usenatbib]{mnras}

\usepackage{newtxtext,newtxmath}

\usepackage[T1]{fontenc}

\DeclareRobustCommand{\VAN}[3]{#2}
\let\VANthebibliography\thebibliography
\def\thebibliography{\DeclareRobustCommand{\VAN}[3]{##3}\VANthebibliography}

\usepackage{graphicx}	
\usepackage{amsmath}	
\usepackage{tabularx}
\usepackage{physics}
\usepackage{float}
\usepackage{orcidlink}
\usepackage[version=4]{mhchem}

\newcommand{\kms}{kms$^{-1}$}
\newcommand{\gravsphere}{\texttt{GravSphere}}

\newcommand{\DMVI}{$\rho_{\rm{DM,VI}}$(150 pc) = (1.4 $\pm$ 0.5) $\times 10^{8}$ M$_{\odot}$ kpc$^{-3}$}
\newcommand{\DMXXIII}{$\rho_{\rm{DM,XXIII}}$(150 pc) = 0.5$\substack{+0.4 \\ -0.3} \times 10^{8}$ M$_{\odot}$ kpc$^{-3}$}

\newcommand{\rhVI}{$r_{\rm{h,VI}} = (489\pm{22})$ pc}
\newcommand{\rhXXIII}{$r_{\rm{h,XXIII}} = 1170 \substack{+95 \\ -94}$ pc}

\newcommand{\MLVI}{(27.1 $\pm$ 8.2) M$_{\odot}$/L$_{\odot}$}
\newcommand{\MLXXIII}{(90.2 $\pm$ 53.9) M$_{\odot}$/L$_{\odot}$}

\newcolumntype{Y}{>{\centering\arraybackslash}X}

\title[Mass Modeling And VI \& And XXIII]{Mass Modeling the Andromeda Dwarf Galaxies: Andromeda VI and Andromeda XXIII}

\author[C.S. Pickett et al.]{Connor S. Pickett\orcidlink{0000-0002-0684-4277}$^{1}$\thanks{E-mail: c.pickett@surrey.ac.uk},
Michelle L.M. Collins\orcidlink{0000-0002-1693-3265}$^{1}$,
R. Michael Rich\orcidlink{0000-0003-0427-8387}$^{2}$,
Justin I. Read\orcidlink{0000-0002-1164-9302}$^{1}$,
Emily J.E. Charles\orcidlink{0000-0002-5886-4202}$^{1}$,
\newauthor
Nicolas Martin\orcidlink{0000-0002-1349-202X}$^{3,4}$,
Scott Chapman\orcidlink{0000-0002-8487-3153}$^{5}$,
Alan McConnachie\orcidlink{0000-0003-4666-6564}$^{6}$,
Alessandro Savino\orcidlink{0000-0002-1445-4877}$^{7}$,
Daniel R. Weisz\orcidlink{0000-0002-6442-6030}$^{7}$
\\
$^{1}$Department of Astronomy, University of Surrey, Stag Hill Campus, Guildford GU2 7XH, UK \\
$^{2}$ Department of Physics and Astronomy, University of California, Los Angeles, PAB, 430 Portola Plaza, LA, CA 90095-1547 \\
$^{3}$Universit\'e de Strasbourg, CNRS, Observatoire astronomique de Strasbourg, UMR 7550, F-67000 Strasbourg, France \\
$^{4}$Max-Planck-Institut f\"{u}r Astronomie, K\"{o}nigstuhl 17, D-69117 Heidelberg, Germany \\
$^{5}$Department of Physics and Atmospheric Science, Dalhousie University, 1453 Lord Dalhousie Drive, Halifax, NS B3H 4R2, Canada \\
$^{6}$NRC Herzberg Astronomy and Astrophysics, 5071 West Saanich Road, Victoria, BC V9E 2E7, Canada\\
$^{7}$Department of Astronomy, University of California Berkeley, Berkeley, CA 94720, USA \\
}

\date{Accepted XXX. Received YYY; in original form ZZZ}

\pubyear{2024}

\begin{document}
\label{firstpage}
\pagerange{\pageref{firstpage}--\pageref{lastpage}}
\maketitle

\begin{abstract}
Accurately mapping the mass profiles of low mass dwarf spheroidal (dSph) galaxies allows us to test predictions made by dark matter (DM) models. To date, such analyses have primarily been performed on Milky Way (MW) satellites. Meanwhile, the Andromeda Galaxy (M31) is home to 35 known dwarf galaxies, yet only two have been successfully mass-modeled so far. A more comprehensive study of Local Group dwarfs is necessary to better understand the nature of dark matter. In this study, we have undertaken a dynamical study of two higher-luminosity Andromeda dwarf galaxies: Andromeda VI (And VI) and Andromeda XXIII (And XXIII). We infer an enclosed mass for And VI of M(r $<$ r$_{h}$) = (4.9 $\pm$ 1.5) $\times$ 10$^{7}$ M$_{\odot}$, corresponding to a mass-to-light ratio of $[M/L]_{r_{\rm{h}}}$ = \MLVI. We infer an enclosed mass for And XXIII of M(r $<$ r$_{h}$) = (3.1 $\pm$ 1.9) $\times$ 10$^{7}$ M$_{\odot}$, corresponding to a mass-to-light ratio of $[M/L]_{r_{\rm{h}}}$ = \MLXXIII. Using the dynamical Jeans modeling tool, \gravsphere, we determine And VI and And XXIII's dark matter density at 150 pc, finding \DMVI and \DMXXIII. Our results make And VI the first mass-modeled M31 satellite to fall into the cuspy regime. And XXIII has a lower density, implying either a more cored central dark matter density, or a lowering of the density through tides, with early quenching times disfavoring core formation via stellar feedback. This adds And XXIII to a growing list of M31 dwarfs with a central density lower than most MW dwarfs and lower than expected for isolated dwarfs in the Standard Cosmology. This could be explained by the M31 dwarfs having experienced stronger tides than their MW counterparts.

\end{abstract}





\section{Introduction} \label{sec:intro}
Dwarf galaxies are the most abundant form of galaxies in the Local Group (LG) \citep{Schechter_1976, Driver_1994, Blanton_2005, Baldry_2008, McNaught_2014}, with more than 100 identified to date within $\sim$2 Mpc from the Milky Way (MW) \citep{McConnachie_2012, Putman_2021}. A specific class of dwarf galaxies, dwarf spheroidals (dSphs), are observed to have higher mass-to-light ([M/L]) ratios ranging from 10 to 1000s. dSphs are distinguished from other dwarf galaxy types (i.e. dwarf irregulars) by their lack of HI gas, elliptical shape, low luminosity ($-13 \leq M_{\rm{V}} \leq -4$), low mass ($M_{\rm{tot}} \sim 10^{7} M_{\odot}$), older stellar population, and smooth distribution of stars \citep{Gallagher_1994, Mateo_1998, Grebel_2001, Grebel_2003, Belokurov_2007, Walker_2007}. Dwarf spheroidals' higher [M/L] ratios make them ideal testing grounds for testing dark matter models \citep{Tolstoy_2009, Simon_2019}, both by comparing their internal DM distribution to models \citep{Read_2018, Hiyashi_2020, Charles_2022, Hiyashi_2023a, Hiyashi_2023b} and by hunting for gamma-ray or X-ray photons from annihilating and/or decaying dark matter \citep{Alvey_2021, Correa_2021, Zoutendijk_2021, Battaglia_2022, Zimmerman_2024}. Technological improvements over the past two decades have made it possible to observe Andromeda (M31) satellites in great detail, to depths that had previously only been achieved for local MW satellites \citep{Koposov_2008, Martin_2013, Drlica_2015,Simon_2019, Drlica_2020}. Surveys, such as the Pan-Andromeda Archaeological Survey (PAndAS), have been instrumental in shedding light on the systems surrounding M31 \citep{McConnachie_2009, McConnachie_2012, McConnachie_2018}.


Measuring parameters of a dwarf galaxy, such as its kinematics and dark matter density, can be done using a technique known as "mass modeling". This analysis has evolved over the decades, from models of star clusters \citep{Eddington_1915} to probing kinematics of local-field dwarfs' inner regions \citep{Rubin_1978}. Increased measurement precision over time has allowed for improved methods for estimating the central mass of these objects \citep{Peebles_1972}, eventually resulting in dynamical mass modeling of stellar tracers to infer the underlying mass distribution \citep{Jeans_1922, BinneyMamon_1982, Read_2017}. Recent calculations use dynamical Jeans modeling \citep{Jeans_1922, GalDyn_Binney_2008, Read_2017, Hiyashi_2020, Collins_2021, Charles_2022, Hiyashi_2023a, Hiyashi_2023b, Vaz_2023, Julio_2024, deLeo_2024}, as well as alternative methods such as those shown in \citet{Pascale_2019} and \citet{Alvarez_2020}, to model the mass distribution of dark matter. In doing so, a dwarf galaxy's mass and velocity distributions can help uncover its dark matter density profile. The \gravsphere ~software (described in detail in \citealt{Read_2017}, \citealt{Read_2019}, \citealt{Genina_2020}, and \citealt{Collins_2021}) breaks the well-known mass-anisotropy degeneracy by fitting for virial shape parameters, yielding a more accurate recovery of the dwarfs' dark matter content. 

Mass modeling of MW dSphs has yielded several key insights. Firstly, all dSphs are dark matter rich, with total DM masses between $10^{7} $M$_{\odot}$ and $10^{8} $M$_{\odot}$ \citep{Mateo_1998}. Early results in \citet{Strigari_2008} show that many, possibly all, of the dwarfs inhabit a similar DM halo to one another. However, this universality has seemed to decrease with the discovery of lower surface brightness dwarfs like Crater II and Antlia II over the last decade \citep{Walker_2009, Collins_2011, Torrealba_2016, Torrealba_2019}. Secondly, when selecting dwarfs in a weak tidal field today, the dwarfs have a central dark matter density that anti-correlates with the duration of their star formation \citep{Read_2019, deLeo_2024}. This is what was predicted by "dark matter heating" models, in which repeated gas inflow/outflow irreversibly causes the inner dark matter halo to expand \citep{Read_2005, DiCintio_2014, Pontzen_2012, Pontzen_2014, Read_2016}. Thirdly, the densest Milky Way dwarfs have been used to place new and tight constraints on alternative DM models, like Self Interacting Dark Matter \citep{Read_2018, Correa_2021}, fermionic dark matter \citep{Dalcanton_2001, DiPaolo_2018, Alvey_2021}, and bosonic -- or wave-like -- dark matter \citep{GonzalezMorales_2017, Hiyashi_2021, Zimmerman_2024}.

The above results alleviate two long-standing tensions in near-field cosmology. The first is the "cusp-core" problem \citep{Flores_1994, Moore_1994, deBlok_2010}. This is a mismatch between DM-only simulations in Lambda CDM predicting dwarfs to have high central dark matter densities, or "cusps," and observed isolated dwarf irregulars favoring lower inner densities, or "cores" \citep{Dubinski_1991,NFW_1996a, NFW_1996b, NFW_1997, Moore_1999, Popolo_2014}. If cores form through DM heating, then galaxies with little star formation should have a higher, cuspier central density, as predicted by pure dark matter structure formation simulations \citep{DiCintio_2014, Onorbe_2015, Libeskind_2020}. This, for the most part, appears to be what is observed for the more tidally isolated  MW dwarfs \citep{Read_2019, deLeo_2024}. The second is the "too big to fail" problem. DM-only simulations predict that more dense dark matter halos should form more dwarf galaxies \citep{Boylan_2011, Boylan_2012, Read_2016}. Observed dwarfs around both the MW and M31 reside within lower density halos, below the expected virial mass threshold of $M_{\text{vir}} \geq 10^{10}$M$_{\odot}$ predicted in simulations \citep{Papastergis_2015}. This problem has therefore been generically expanded to all MW-sized galaxies by \citet{RodriguezPuebla_2013} and to M31 in \citet{Tollerud_2014}. 

However, there is mounting evidence that M31's dwarfs are distinct from the MW's. An over-reliance on the MW dwarfs could therefore bias our understanding of low-mass galaxy formation or the nature of dark matter. \citet{Collins_2011} showed that Andromeda dwarf galaxies ($M_V < -7.9$) had different central masses and "colder" velocity dispersions than their MW counterparts. This caused a scatter within the "Universal" mass profile and the mean rotation curve derived in \citet{McGaugh_2007}. \citet{Collins_2014} confirmed these findings by doubling the sample size used for the comparison. A Universal mass profile could not be fit to both Milky Way and M31 dwarfs simultaneously. Removal of three extreme, low-density dwarfs (Andromeda XIX, XXI, and XXV) led to a stronger agreement between the two host systems. These outliers, however, contributed to the notion that physical processes, such as tides, have played an important role in the lifetimes of M31 satellites.

At a distance of $\sim$800 kpc \citep{Savino_2022}, M31 halo dwarfs can be resolved into stars whose individual spectra can be used in mass modeling tools such as \gravsphere. However, this method has only successfully been completed for two M31 dwarf galaxies: Andromeda XXI and Andromeda XXV so far \citep{Collins_2021, Charles_2022}. Those works found that both of these dwarfs appear to have a low central dark matter density, consistent with a dark matter core rather than a cusp. And XXI had a central dark matter density at 150 pc of $\rho_{\text{DM,XXI}}(150) = 2.6 \substack{+2.4 \\ -1.5} \times 10^{7}$ M$_{\odot}$ kpc$^{-3}$, while And XXV has a $\rho_{\text{DM,XXV}}(150) = 2.3 \substack{+1.4 \\ -1.1} \times 10^{7}$ M$_{\odot}$ kpc$^{-3}$. Following analysis similar to \citet{Read_2016}, \citet{ReadErkal_2019}, and \citet{Weisz_2019}, \citet[C23 hereafter]{Charles_2022} show that the star formation history (SFH) of both And XXI and And XXV are too short for them to have undergone significant DM heating. Thus, their low density is either a result of tidal stripping and shocking by M31 \citep{Read_2006}, or it is a problem for the standard dark matter model \citep{Spergel_2000, Dave_2001, Vogelsberger_2012, Rocha_2013, Dooley_2016, Kim_2016, Read_2018, Correa_2021}.

Two MW dwarfs have a low central density similar to those of And XXI and XXV, as well as too little star formation to have undergone significant DM heating: Crater II \citep{Torrealba_2016} and Antlia II \citep{Torrealba_2019}. In both cases, their orbits around the MW are known from measurements of their line-of-sight velocities and proper motion, supporting the hypothesis that their low central densities owe to tides. \citep{Fu_2019, Borukhovetskaya_2022, Ji_2021, Hammer_2024, Wang_2024}. However, Crater II and Antlia II are both outliers from the population of MW dwarfs, whereas And XXI and And XXV are simply the first two M31 dwarfs that have been looked at. It is therefore important to make similar measurements of all the M31 dwarfs to more definitively assess whether or not DM heating and tides can explain the whole population.

In this paper, we mass model two more Andromeda dwarf galaxies: Andromeda VI (And VI) and Andromeda XXIII (And XXIII), chosen due to their similarity to the mass-modeled systems of Andromeda XXI \citep{Collins_2021} and Andromeda XXV (C23). And VI is a similarly-sized, higher-luminosity dwarf as Andromeda XXI, while And XXIII resembles the size and luminosity of Andromeda XXV. Thus, initial analysis parameters for this project could be informed by the available literature on other M31 dwarfs. Based on photometry and previous dynamics, And VI is likely to be a typical, isolated dwarf spheroidal at a large separation from the center of M31, with a de-projected distance of $D_{\text{M31,VI}} = 281.6 \substack{+8.6 \\ -7.1}$ kpc \citep{Martin_2016, Savino_2022}. The system falls also within the expected dark matter density for its size. Conversely, And XXIII is likely to be an outlier, similar to recently-modeled dwarfs \citep{Collins_2021, Charles_2022}. This system is closer to its host, at a de-projected distance of $D_{\text{M31,XXIII}} = 128.1 \substack{+10 \\ -4.9}$ kpc \citep{Martin_2016, Savino_2022}, and may have had a more turbulent history than And VI. We mass model both dwarfs to measure their central DM densities. We also compare their observed densities at the present epoch to those we would anticipate from their respective star formation histories.

This paper is organized as follows: In $\S$ \ref{sec:observations}, we outline the photometric and spectroscopic analysis of And VI and And XXIII. This section also describes the selection of member stars from these data. In $\S$ \ref{sec:analysis}, we describe the kinematic analysis of the dwarfs and their mass modeling through \gravsphere. We discuss these results in $\S$ \ref{sec:discussion}, where we also compare these M31 satellites with others in the Local Group. Finally, we conclude this work in $\S$ \ref{sec:conclusions}.


\section{Observations} \label{sec:observations}
Andromeda VI and Andromeda XXIII were analyzed using existing data. Photometric observations of And VI were collected using the \textit{Subaru} Suprime-Cam from August 2009 and are described in detail in \citet{McConnachie_2007} and \citet{Collins_2011}. Photometric data of Andromeda XXIII was taken via the Pan-Andromeda Archaeological Survey (PAndAS) from 2008 to 2011 \citep{McConnachie_2009, McConnachie_2012, McConnachie_2018}. Andromeda VI sits at a distance to M31 of $D_{\text{M31}} =$ 281.6$\substack{+8.6 \\ -7.1}$ kpc, outside of the PAndAS footprint, and was therefore not observed in the survey \citep{Savino_2022}. Both dwarfs were spectroscopically observed with the Deep Extragalactic Imaging Multi-Object Spectrograph (DEIMOS) on the Keck II telescope in Hawaii \citep{Faber_2003, McConnachie_2012}. Spectroscopic data for both dSphs is further described in \cite[C13 hereafter]{Collins_2013}, while data reduction of the spectroscopy can be found in \cite{Ibata_2011}. A brief description of each dataset is given below.

\subsection{PAndAS Survey} \label{subsec:PANDAS}
PAndAS collected observations between 2008 and 2011 \citep{McConnachie_2009, McConnachie_2012, McConnachie_2018}. The survey used Mega-Cam, mounted on the Canada-France-Hawaii Telescope (CFHT), on Mauna Kea, which allowed for a 400 deg$^{2}$ view of the M31 halo out to 150 kpc. The goal of the survey was to observe red giant branch (RGB) stars at a distance of  $\sim$780-900~kpc \citep{Conn_2012}. PAndAS observations had a signal-to-noise ratio (SNR) of $\sim$10 at magnitudes of \textit{g} = 25.5 and \textit{i} = 24.5 \citep{McConnachie_2009}. These magnitudes were ideal for detecting dwarf galaxies, which were characterized as overdensities of old faint stellar populations within the survey's view. 

The collected data were reduced using the process described in \citet{Ibata_2014} and \citet{McConnachie_2018}, primarily using the Cambridge Astronomical Survey Unit (CASU) pipeline described in \citet{Irwin_2004}. Sources were identified on the image within a detection threshold of 10$\sigma$ on a chip-by-chip basis with respect to astrometry from \textit{Gaia} DR1 \citep{GaiaDR2_2018a, GaiaDR2_2018b}, reducing zero-point survey uncertainties. The calibrated images were then recombined through the CASU software. Summed \textit{g + i}-band images were then re-run with a 3$\sigma$ threshold, with any returned sources being photometrically remeasured. Each band was then classified into point sources, extended objects, or noise. The final output was then calibrated onto the Pan-STARRS photometric system \citep{Flewelling_2020}. This resulted in photometric uncertainties better than 0.1 magnitude at \textit{g} $\simeq$ 25 and \textit{i} $\simeq$ 24 \citep{McConnachie_2018}.

Using PAndAS data, Andromeda XXIII was measured to have an absolute $V$-band magnitude of $M_{\rm{V,XXIII}} = -9.8 \pm{0.2}$, a luminosity of $L_{\rm{V,XXIII}}$ = (6.3 $\pm$ 0.1) $\times ~10^{5}$ L$_{\odot}$, and a half-light radius of \rhXXIII ~\citep{McConnachie_2012, Martin_2016, Savino_2022}. With a measured MW-to-M31 distance of $D_{\odot,\text{M31}} = 776.2 \substack{+22\\ -21}$ kpc, \citet{Savino_2022} also measured And XXIII, through HST-based RR Lyrae observations, to have a distance from M31 of $D_{\text{M31,XXIII}} = 128.1 \substack{+10\\ -4.9}$ kpc and a distance to the Milky Way of $D_{\text{MW,XXIII}} = 745 \substack{+24 \\ -25}$ kpc.


\subsection{Subaru Suprime-Cam Photometry} \label{subsec:subaru}
Photometric data were collected by the \textit{Subaru} Suprime-Cam on 21-22 August 2009 in the Cousins V and i$_{\rm{c}}$ filters. The average seeing was $\sim$1 arcsecond. Observations were taken in a single field of 5 $\times$ 440s in the V-band and 20 $\times$ 240s in the i$_{\rm{c}}$-band. These data were run through a CASU pipeline which debiased, flat-fielded, trimmed, and gain-corrected the images \citep{Irwin_2001}. A catalogue was generated with morphologically-separated stellar objects from non-stellar and noise-like objects. These data were also extinction corrected through dust maps in \citet{Schlegel_1998}.

\begin{figure}
    \centering
    \includegraphics[width=\columnwidth]{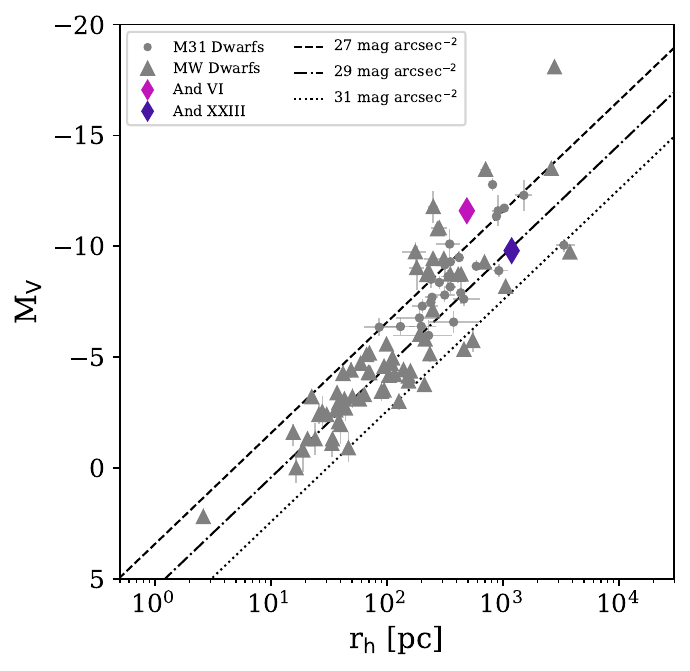}
        \caption{Half-light radius versus absolute V-band magnitude of Local Group dwarf galaxies. Milky Way dwarfs are denoted by gray triangles, while M31 dwarfs are denoted by gray circles. Andromeda VI and Andromeda XXIII are represented by a magenta diamond and a dark purple diamond, respectively. The black lines represent relations of surface brightness. Note that some uncertainties are too small to be seen but are represented by the size of the markers. The data shown in gray are from \citet{Savino_2022} and \citet{Pace_2024}.}
    \label{fig:rvMagV}
\end{figure}

These observations were used as a follow-up to the discovery of Andromeda VI in 1999 by \cite{Armandroff_1999}, measuring an absolute magnitude of And VI to be  $M_{\rm{V,VI}} = -11.3 \pm{0.2}$ and a half-light radius of $r_{\rm{h}} = (524\pm{49})$ pc \citep{McConnachie_2012}. These values were later updated in \citet{Savino_2022} to $M_{\rm{V,VI}} = -11.6 \pm{0.2}$ and \rhVI, with a distance to M31 of $D_{\text{M31,VI}} = 281.6 \substack{+ 8.6\\ -7.1}$ kpc and a distance to the Milky Way to be $D_{\text{MW,VI}} = (831.8 \pm{23})$ kpc. The $M_{\rm{V}}$-r$_{\rm{h}}$ relation for both Andromeda VI and Andromeda XXIII can be seen in Figure \ref{fig:rvMagV}.

 Using the morphological shape parameters defined in the CASU pipeline, we removed any objects likely to be background galaxies or noise, while keeping those objects classified as stars. Stars within 2 half-light radii of the dwarf's assumed center are tagged as such and are used to make a clear color magnitude diagram (CMD). Retaining these objects visually clarified the RGB overdensities of both dwarfs, making it easier to overlay an isochrone, as described in $\S$ \ref{sec:analysis}. All objects, regardless of radius from the assumed center, are used in the following analysis.

\subsection{DEIMOS Spectroscopy} \label{subsec:DEIMOS}
Spectroscopic observations of Andromeda VI and Andromeda XXIII were completed using the Deep Extragalactic Imaging Multi-Object Spectrograph (DEIMOS), mounted on the Keck II telescope in Mauna Kea, Hawaii \citep{Faber_2003}. Around 150 objects were included in a given pointing via the spectrograph's multiobject mode, providing large amounts of data across a $16 \times 8$ arcmin$^{2}$ field of view (FoV). At the distances of Andromeda dwarf galaxies, this FoV translates to $\sim$3.5kpc $\times$ 1.3kpc. With an observed half-light radius of \rhVI, And VI can fit comfortably within the spectrograph's FoV. Conversely, the FoV can only capture And XXIII to one half-light radius of \rhXXIII ~\citep{McConnachie_2012, Martin_2016}. DEIMOS can provide observations in the optical to near-infrared spectrum, between the wavelength range of $\lambda =$ 400 \AA ~- 11,000 \AA, within which calcium triplet lines (\ce{Ca II}) can be identified at $\sim$8500 \AA. These strong calcium absorption features are used to determine the velocities and metallicities of observed stars within a given galaxy \citep{Wilkinson_2004, Collins_2010, Starkenburg_2010, Tollerud_2012, Collins_2021, Charles_2022}. All masks were collected with the same 1200 line mm$^{-1}$ grating at a resolution of 1.3 \AA, with data collected between 5800-9800 \AA, centered at $\lambda \sim$ 7800 \AA ~to resolve \ce{Ca} II lines. These were taken between September 2011 (described in C13) and October 2013 (during a PAndAS follow-up survey, called Z-PAndAS \citep{Collins_2017}) as part of our ongoing study of the Andromeda dwarf galaxy system. Full details of these observations can be found in C13 and \cite{Collins_2020}, with key reduction steps being briefly described in this work. Andromeda VI spectra were comprised of one mask, taken in 2011. This mask was split into 3 $\times$ 1200 second exposures, resulting in a total exposure time of 3600s. Alongside the 2011 mask, Andromeda XXIII spectroscopy also comprised of a 2013 DEIMOS observation. The 2013 mask was split into 6 $\times$ 1200 second exposures, resulting in a total time of 7200s. The average seeing per mask was $0.6^{\prime\prime}-0.8^{\prime\prime}$. A reasonable mean SNR of $\sim$5 per pixel was achieved, with 102 observed stars for And VI and 115 observed stars for And XXIII. Duplicate stars have been removed from And XXIII observations by visually inspecting both masks to find stars with the same right ascension and declination. The star that best matches the systemic velocity of And XXIII is kept. If the measured velocities are indistinguishable, the star with the lowest velocity error is kept. Eight duplicates have been identified between both masks, resulting in a total of 107 unique observed stars for Andromeda XXIII.

Data reduction was completed through a custom pipeline, described in detail in \citet{Ibata_2011} and C13. This pipeline identifies and corrects for cosmic rays, scattered light, illumination, slit function, and fringing. Corrections of pixel-to-pixel variations are also made via flat-fielding, while wavelength calibrations of each pixel are completed using arc-lamp exposures. Sky subtractions, from a small spatial region around a given target, were taken from the two-dimensional spectra prior to spectrum extraction without resampling. The velocities of each star were measured using \ce{Ca} II absorption features, with each velocity being determined through a Markov chain Monte Carlo (MCMC) procedure. This was done by cross-correlating the non-resampled data with a \ce{Ca}II spectrum template. The resulting values were the most likely velocities for each star. A corresponding uncertainty value for each velocity was output and accounted for all uncertainties for each pixel. The inherent uncertainty for DEIMOS was taken to be 3.2 \kms, as shown in \citet{Collins_2020}. The velocities were then corrected to the heliocentric frame.

Slitmask misalignments, which can cause velocity shifts up to 15 \kms, were corrected by comparing telluric absorption line in the data to atmospheric models. Doing so shifted each spectra to the correct frame \citep{Ibata_2011, Collins_2013}. Velocity uncertainty gradients were not found in either mask, and any stars present in both masks were compared to ensure proper uncertainty values. The final velocity uncertainty is a combination of MCMC posterior distribution and inherent DEIMOS uncertainty, as well as uncertainty of the reduction pipeline at 3.2 \kms. This value is chosen to mirror \citet{Collins_2020} and C23 for our analysis, however it should be noted that \citet{Martin_2014} uses 3.4 \kms for similar targets.

DEIMOS data are matched with available photometry, within a maximum separation of 1 arcsecond, to their positions using the \texttt{astropy} Python package \citep{Astropy_2013, Astropy_2018, Astropy_2022}. Membership to the dwarf is determined by aligning an old, metal-poor, alpha-enhanced isochrone to the stars through a \texttt{Padova} stellar evolution model \citep{ODonnell_1994, Kroupa_2001, Kroupa_2002, Groenewegen_2006, Marigo_2008, Marigo_2013,Chen_2014, Tang_2014, Chen_2015, Rosenfield_2016, Pastorelli_2019, Trabucchi_2019, Bohlin_2020, Pastorelli_2020, Trabucchi_2021}. And VI is matched with an isochrone of age $10$ Gyr, metallicity of [Fe/H] = -1.5 dex, and extinction of $A_{\rm{v}} = 1.2$. And XXIII is matched with an isochrone of age $10$ Gyr, metallicity [Fe/H] = -1.8 dex, and extinction of $A_{\rm{v}} = 0.33$. These values can also be seen in Table \ref{tab:isochrone_params}, alongside further explanation of CMD membership in Section \ref{subsubsec:CMD}.


\section{Mass-Modeling Methods} \label{sec:massmodelmethods}

We use a dynamical mass-modeling tool, \gravsphere\footnote{\url{https://github.com/justinread/gravsphere}}, to measure the dark matter density profiles of Andromeda VI and Andromeda XXIII. This software is described in detail in \citet{Read_2017}, \citet{Read_2018}, \citet{Read_2021}, \citet{Genina_2020}, and \citet[C21 hereafter]{Collins_2021}. \gravsphere ~solves the spherical Jeans equation \citep{Jeans_1922} for a set of member stars, or `tracers', with radial velocity measurements. In doing so, \gravsphere ~can then solve for the dark matter density, $\rho(r)$, and the velocity anisotropy profile, $\beta(r)$. The Jeans equation, as derived in \citet{vanderMarel_1994} and \citet{Mamon_2005}, is as follows:

\begin{equation}\label{eq:Jeans}
    \sigma_{\rm{LOS}}^{2}(R) = \frac{2}{\Sigma_{*} (R)} \int_{R}^{\infty}\Bigg(1 - \beta(r) \frac{R^{2}}{r^{2}}\Bigg)  \frac{\nu(r) \sigma_{r}^{2}(r) r}{\sqrt{r^{2} - R^{2}}}\text{d} r
\end{equation}
where $\sigma_{\rm{LOS}}$ is the line-of-sight velocity of the member stars, $\Sigma_{\rm{*}}(R)$ is the surface brightness profile at radius $R$ from the galaxy's center, $\nu(r)$ is the spherically averaged member star density as a function of $r$, and $\beta(r)$ is the velocity anisotropy:

\begin{equation}\label{eq:BetaAnisotropy}
    \beta(r) \equiv 1 - \frac{\sigma_{\rm{t}}^{2}}{\sigma_{\rm{r}}^{2}}
\end{equation}
where $\sigma_{\rm{t}}$ is the tangential velocity dispersion profile and $\sigma_{\rm{r}}$ is the radial velocity dispersion profile. The radial velocity dispersion profile can be expanded as:

\begin{equation} \label{eq:RadialProfile}
    \sigma_{\rm{r}}^{2} = \frac{1}{v(r)g(r)} \int^{\infty}_{\rm{r}} \frac{GM(\tilde{r})v(\tilde{r})}{\tilde{r}^{2}}g(\tilde{r})\text{d}r
\end{equation}
where

\begin{equation} \label{eq:g_of_r}
    g(r) = \exp \Bigg(2 \int \frac{\beta(r)}{r}\text{d}r\Bigg)
\end{equation}
and $M(r)$ is the total cumulative mass as a function of radius $r$. $\beta$ is the anisotropy parameter, which describes how anisotropic the stellar orbital velocities are: $\beta$ = 0 represents an isotropic distribution, $\beta$ = 1 represents a fully radial distribution, and $\beta$ = -$\infty$ represents a fully tangential distribution. A symmetrized version of $\beta$ is used by \gravsphere ~to avoid infinite values when solving for velocity anisotropy:

\begin{equation} \label{ew:sym_beta}
    \tilde{\beta} = \frac{\sigma_{\rm{r}} - \sigma_{\rm{t}}}{\sigma_{\rm{r}} + \sigma_{\rm{t}}} = \frac{\beta}{2 - \beta}
\end{equation}
where $\tilde{\beta}$ = -1 corresponds to a fully tangential anisotropy, $\tilde{\beta}$ = 0 corresponds to isotropy, and $\tilde{\beta}$ = 1 corresponds to a fully radial anisotropy \citep{Read_2017}.

For the DM mass profile, we use the \texttt{CoreNFWTides} model \citep{Read_2018, Collins_2021}. This, at its heart, has a cusped Navarro-Frenk-White (NFW) profile \citep{NFW_1996b}. The NFW profile contains a virial mass, $M_{\rm{200}}$, and a concentration parameter, $c_{\rm{200}}$. \texttt{CoreNFWTides} also adds four new parameters: $n$, $r_{\rm{c}}$, $r_{\rm{t}}$, and $\delta$. $n$ controls how `cusped' or `cored' a dark matter profile is within $r_{\rm{c}}$. $n$ = -1 corresponds to a steep cusp of $r^{-2}$, $n$ = 0 corresponds to a cusp of $r^{-1}$, and $n$ = 1 corresponds to a flat core. $r_{\rm{t}}$ and $\delta$ help model tidal forces that occur when a more massive host galaxy strips material from the outer edges of a dwarf, where the tidal radius $r_{\rm{t}}$ represents the point beyond which density decreases as $r^{-\delta}$.

Both And VI and And XXIII have some mild ellipticity, while \gravsphere ~assumes they are spherical. \gravsphere ~has been rigorously tested on mock data of triaxial dwarfs with a similar number of available stellar velocities as And VI and And XXIII \citep{Read_2017, Read_2018, Genina_2020, Collins_2021, Nguyen_2025}. These tests show that, despite each galaxy's asymmetry ($\epsilon_{\rm{VI}} = 0.41 \pm 0.03$; $\epsilon_{\rm{XXIII}} = 0.40 \pm 0.05$), any present bias is smaller than our quoted formal uncertainties. Furthermore, \citet{deLeo_2024} has shown through mock data tests that the software also works in the presence of extreme tides. Recently, updates were made to \gravsphere ~to eliminate binning bias towards cusped profiles when using low numbers of member stars or large velocity uncertainties. The implementation of \texttt{Binulator}, which fits a Gaussian probability distribution to each bin, provides stronger estimates of the mean, variance, kurtosis, and uncertainties. The results are then used as inputs for \gravsphere. This process applies to small data sets, as shown in C23 where only 49 member stars are used to analyze a dwarf galaxy. Further testing has been detailed in Appendix A of C21.

Updates have also been made to \gravsphere ~to more tightly constrain priors on the MCMC routine when fitting for dark matter density. These updates are outlined in Table \ref{tab:GravSphereUpdatedPriors} and show improvements in dark matter density measurements and uncertainties, specifically when compared to previous versions of \gravsphere. For example, it is now less likely for uncertainties, within 2$\sigma$, to indicate a cored profile when a cusp is expected. The priors used in \gravsphere ~v1.5 still disfavor perfect cusps and cores, as described in \citet{Read_2017} for ~v1.0. This is because the correct solution resides on the edge of the hypervolume of solution space, which is large due to the $\rho - \beta$ degeneracy. Correct models are contained with the results but are rare, presenting instead a constrained range of estimations rather than a singular, definitive profile. Changes were also made to $n$, which sets the shape of the dark matter density profile. We have previously used coreNFWtides priors on $n$ symmetrized on $-1 < n < 1$ \citep{Collins_2021}. This symmetrizes the solution around an NFW cusp in the absence of data. However, it also allows solutions steeper than NFW, which are not expected in the standard cosmology. Hence, in \gravsphere ~v1.5 we favor a more restrictive prior in the range $0 < n < 1$. We explore the impact of this choice for Andromeda VI, shown in Figure \ref{fig:And6_prior_den_comp} in Appendix \ref{subsec:append_plots}, where we run \gravsphere ~again with the broader $-1 < n < 1$ prior. We find that there is negligible difference between the two profiles, with the only notable change being the estimation of slightly lower densities at extreme radii in v1.5 when compared to v1.0. Both profiles converge around the half-light radius and are virtually indistinguishable from one another at 150 pc. We would anticipate a similar result for Andromeda XXIII. Remaining \gravsphere ~parameters used for Andromeda VI and Andromeda XXIII are detailed in $\S$ \ref{subsec:binulator_gravsphere}.

\begin{table}
\centering
\renewcommand{\arraystretch}{1.4}
    \caption{Updates to \texttt{GravSphere} priors. The left column shows priors used in C21 and C23. The right column is used in this work.}
    \label{tab:GravSphereUpdatedPriors}
    \begin{tabularx}{\columnwidth}{>{\centering\arraybackslash}X | >{\centering\arraybackslash}X}
    \texttt{GravSphere} v1.0 & \texttt{GravSphere} v1.5 \\ \hline
    7.5 $< \log(M_{\rm{200}}) <$ 11.5 & 9.0 $< \log(M_{\rm{200}}) <$ 10.0 \\
    -1.0 $< n <$ 1.0 & 0 $< n <$ 1.0 \\
    3.01 $< \delta <$ 5.0 & 3.01 $< \delta <$ 4.0 \\
    \end{tabularx}
\end{table}


\section{Analysis and Results} \label{sec:analysis}
In this section, we present our measurements of the dynamics of both Andromeda VI and Andromeda XXIII, the characteristics of which can be seen in Table \ref{tab:dwarf_chars}. The membership determination process described in C13 follows a maximum-likelihood (ML) estimation, similar to that outlined in \citet{Martin_2007}. The ML approach utilized sigma clipping when identifying likely members, where the most optimal values were trimmed to within $\pm{3\sigma}$ of an accepted `mean' value. The method in our analysis has since been updated to a Bayesian approach and can be described, in detail, in \citet{Collins_2020}, C21, and C23. In Bayesian analysis, each star is assigned a membership probability based on two criteria: (1) position on a color magnitude diagram of the dwarf ($P_{\rm{CMD}}$) and (2) velocity ($P_{\rm{vel}}$). Both of these probability assignments are described in greater detail below. Previous works use a third probability weight of radial distance, however we opt to exclude this factor. It was shown in C13 that this is most useful for dwarf galaxies where the systemic velocity is similar to that of the contaminant population (typically the MW halo). The data provided are relatively clean in this regard, and the impact of radial distance probabilities is thus negligible in our analysis.

\begin{table*}[!t]
    \centering
    \renewcommand{\arraystretch}{1.25}
    \caption{Properties of Andromeda VI and Andromeda XXIII. Sources: $^{a}$\citet{McConnachie_2012}, $^{b}$\citet{Savino_2022}, $^{c}$This work.}
    \label{tab:dwarf_chars}
    \begin{tabularx}{\textwidth}{X X l}
        \\ \hline
        Property & And VI & And XXIII \\ \hline
        $^{a}$$\alpha$, $\delta$ (J2000) &  $23^{h}51^{m}46^{s}.3, +24^{\circ}34'57''$ & $01^{h}29^{m}21.8^{s}, +38^{\circ}43'26''$ \\
        
        $^{b}$m$_{\rm{V}}$& 24.6 $\pm$ 0.06  & 14.6 $\pm$ 0.2\\
        
        $^{b}$M$_{\rm{V}}$&  -11.6 $\pm$ 0.2 & -9.8$\pm$ 0.2\\
        
        $^{b}$D$_{\rm{\odot}}$ (kpc)& 831.8 $\pm$ 23 & 745.0 $\substack{+24 \\ -25}$\\

        $^{b}$D$_{\text{M31}}$ (kpc)& 281.6 $\substack{+8.6 \\ -7.1}$  & 128.1 $\substack{+10 \\ -4.9}$\\

        $^{a}$$\epsilon$ & 0.41 $\pm$ 0.03 & 0.40 $\pm$ 0.05\\
        
        $^{a}$$\theta$ ($^{\circ}$) & 163 $\pm$ 3 & 138 $\pm$ 5\\
        
        $^{a}$r$_{\rm{h}}$ (arcmin)&  2.3 $\pm$ 0.2 & 4.6 $\pm$ 0.2 \\
        
        $^{b}$r$_{\rm{h}}$ (pc)& 489 $\pm$ 22 & 1170 $\substack{+95 \\ -94}$ \\
        
        $^{c}$v (\kms)& -341.6 $\pm$ 1.7 & -237.7 $\pm$ 1.4 \\
        
        $^{c}$$\sigma_{\rm{v}}$ (\kms)&  13.2 $\substack{+1.4 \\ -1.3}$ & 6.8 $\substack{+1.5\\ -1.3}$\\

         $^{c}$L (r $<$ r$_{\rm{h}}$) ($\times 10^{5}$ L$_{\rm{\odot}}$)& 18.2 $\pm$ 0.4 & 3.5 $\pm$ 0.1 \\
        
        $^{c}$M (r $<$ r$_{\rm{h}}$) ($\times 10^{7}$ M$_{\odot}$)& 4.9 $\pm$ 1.5 & 3.1 $\pm$ 1.9\\
        
        $^{c}$$[$M/L$]$$_{\rm{r}_{\rm{h}}}$ (M$_{\odot}$/L$_{\odot}$)&  27.1 $\pm$ 8.2 & 90.2 $\pm$ 53.9 \\
        
        $^{a}$$[$Fe/H$]$ (dex)&  -1.5 $\pm$ 0.1 & -1.8 $\pm$ 0.1 \\
        
        $^{c}$$\rho_{\rm{DM}}$(150 pc) ($\times$10$^{8}$ M$_{\odot}$ kpc$^{-3}$)& 1.4 $\pm$ 0.5 & 0.5 $\substack{+0.4 \\ -0.3}$\\ 
        
        $^{c}$M$_{200}$ ($\times 10^{9}$ M$_{\odot}$)&  12.7 $\pm$ 4.7 & 3.9 $\pm$ 1.6 \\ \hline
         
    \end{tabularx}
\end{table*}

\subsection{Membership Probability}\label{subsec:membership_prob}
The photometric data from \textit{Subaru} and PAndAS (for And VI and And XXIII, respectively) are used to create a CMD. The data from \textit{Subaru} are plotted in the \textit{V}- and \textit{i}-bands, while the data from PAndAS are plotted in the \textit{g}- and \textit{i}-bands. The CMD for each galaxy can be seen on the right-hand plot of Figures \ref{fig:And6all} and \ref{fig:And23all}. In this section, we provide a brief explanation of each probability cut applied to the data.

\subsubsection{CMD Probability} \label{subsubsec:CMD}

\begin{table}
\renewcommand{\arraystretch}{1.25}
    \centering
    \caption{Isochrone parameters for And VI and And XXIII, where $A_{\rm{v}}$ is the extinction coefficient. A [Fe/H] value of -1.5 dex more centrally fit our And VI data, hence the difference from a value of -1.3 dex seen in the literature.}
    \label{tab:isochrone_params}
    \begin{tabularx}{\columnwidth}{X | Y | Y | Y}
    \textbf{Galaxy} & $A_{\rm{v}}$ & Age (yr) & ([Fe/H]) \\ \hline
    And VI & 1.2 & 1 x 10$^{10}$ & -1.5 \\
    And XXIII & 0.55 & 1 x 10$^{10}$ & -1.8
    \end{tabularx}
\end{table}

Stars are determined to lie on a given dwarf's RGB, via alignment to the generated \texttt{Padova} isochrone (see Sec. \ref{subsec:DEIMOS}), through the following equation:

\begin{equation} \label{eq:distance_equation}
    R = \sqrt{(x_{2} - x_{1})^{2} + (y_{2} + y_{1})^{2}}
\end{equation}
where (x$_{1}$, y$_{1}$) are the x-y coordinates of the isochrone (i.e. the color and the magnitude) and (x$_{2}$, y$_{2}$) are the x-y coordinates of the matched stars. Isochrone data is interpolated onto a fine grid using the \texttt{scipy.interpolate} package before the distance from the isochrone is measured \citep{Scipy_2020}. The distance, $R_{\rm{min}}$, for each individual star is calculated and applied to a Gaussian probability:

\begin{equation} \label{eq:Gauss_CMD}
    P_{\rm{CMD}} = \exp \Bigg(-\frac{1}{2} \Bigg(\frac{R_{\rm{min}}^{2}}{\eta _{\rm{CMD}}^{2}}\Bigg)\Bigg)
\end{equation}
where $\eta_{\rm{CMD}}$ is a scaling parameter that accounts for the scatter of stars around the isochrone. We use a value of $\eta_{\rm{CMD}}$ = 0.1 for both Andromeda VI and Andromeda XXIII. $\eta_{\rm{CMD}}$ allows for a reasonable width for the RGB as opposed to considering a single stellar population.

\subsubsection{Velocity Probability} \label{subsubsec:velocity}
\begin{figure}
    \centering
    \includegraphics[width = \columnwidth]{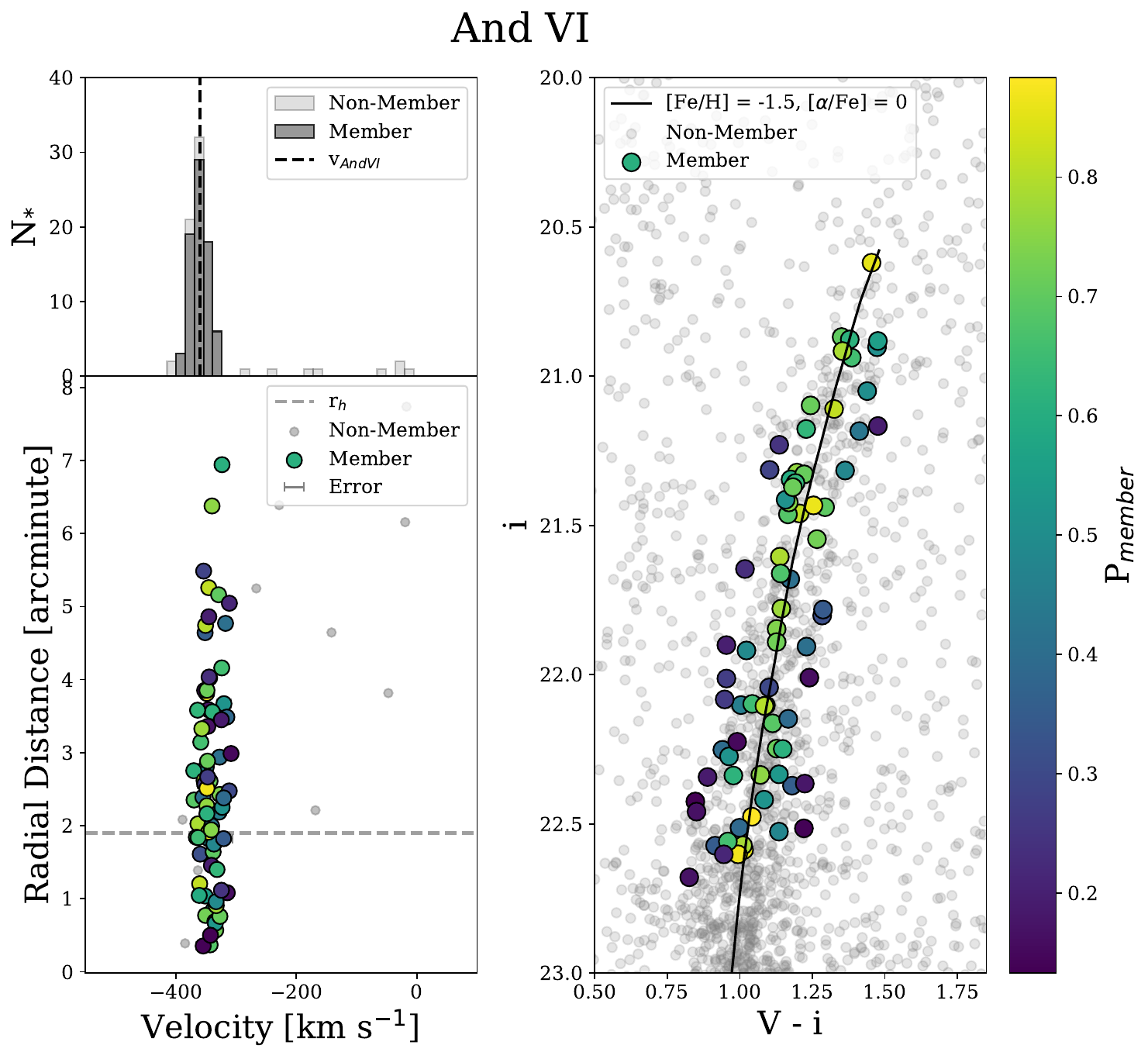}
    \caption{\textbf{Top Left}: Histogram of number of stars split into 40 velocity bins of 15 \kms ~each. The light gray bars represent the non-member velocities while the dark gray bars represent member star velocities. The vertical dashed black bar is the systemic velocity of And VI from C13. \textbf{Bottom Left}: Velocity versus radial distance from measured center of And VI. The points are shaded depending on their respective probabilities. Each point's uncertainty is too small to be seen relative to each point but is adequately represented by the size of each marker. The dashed light gray horizontal line is the measured half-light radius of And VI from \citet{Savino_2022}. \textbf{Right}: CMD of the RGB overdensity of And VI. The light gray points are non-member stars, as determined by our probability calculations. The colored points represent member stars, shaded according to their respective probability. The RGB is overlaid with the best-fit-by-eye isochrone of parameters [Fe/H] = -1.5 dex, [$\alpha$/Fe] = 0, and age = 10 Gyr.}
    \label{fig:And6all}
\end{figure}

\begin{figure}
    \centering
    \includegraphics[width = \columnwidth]{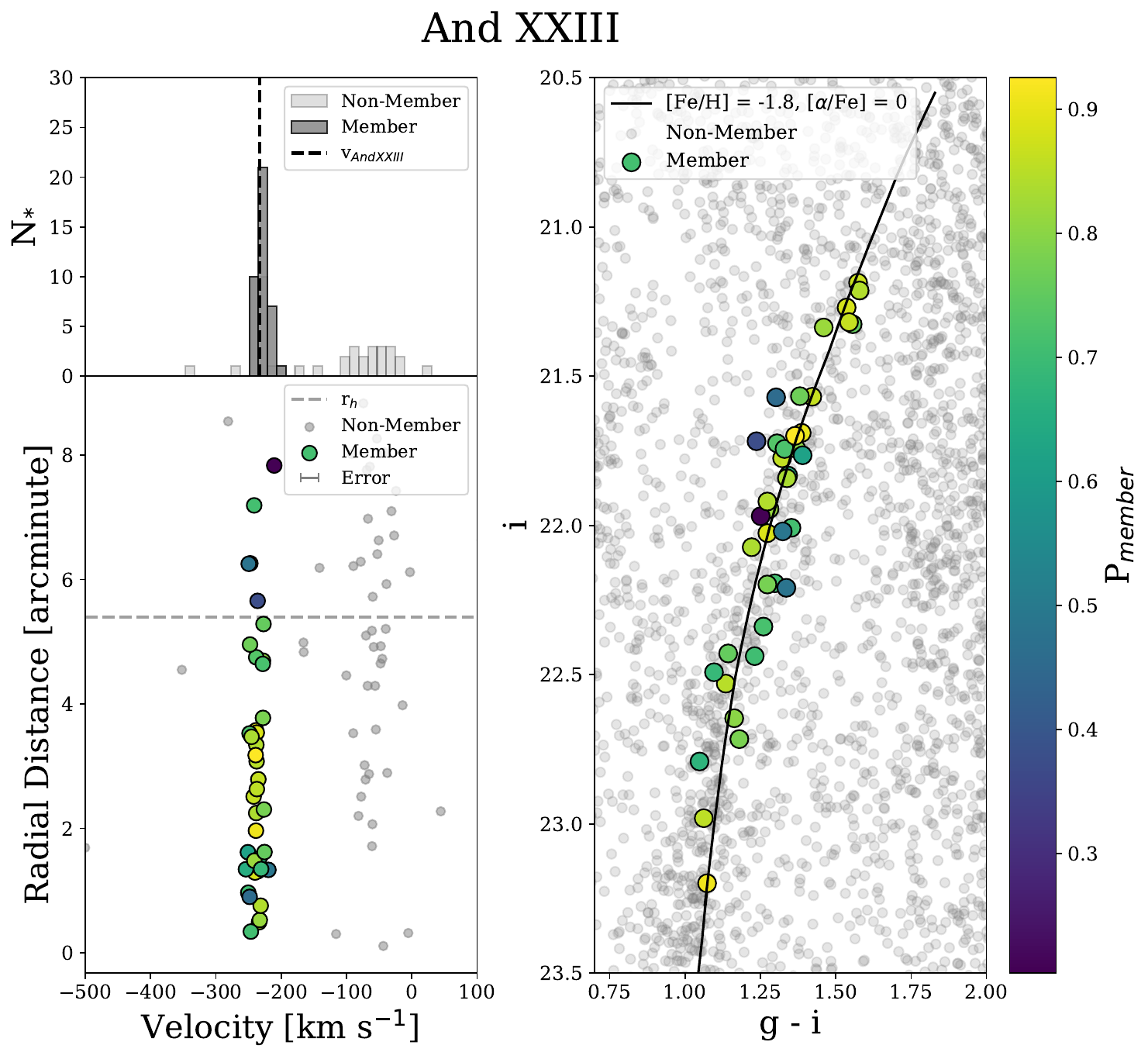}
    \caption{Same as Figure \ref{fig:And6all} but for And XXIII. The best-fit-by-eye isochrone overlaid on the RGB has the parameters [Fe/H] = -1.8 dex, [$\alpha$/Fe] = 0, and age = 10 Gyr.}
    \label{fig:And23all}
\end{figure}

Figures \ref{fig:And6all} and \ref{fig:And23all} show a clear population at $v \approx -350$ \kms and $v \approx -240$ \kms, for And VI and XXIII respectively. However, there also exists populations at higher velocities, which is indicative of contamination. Milky Way contamination is typically centered around 0 \kms but can extend to -200 \kms, while M31 contamination is centered around -300 \kms with a broad dispersion \citep{Chapman_2006}. Dynamics of each population can be used to delineate membership between these velocity groups. The probability of each star as belonging to our desired dwarf population, $P_{\rm{vel}}$, is determined through an MCMC procedure, outlined in \citet{Goodman_2010}, using the \texttt{emcee} Python package \citep{Foreman-Mackey_2013}. Analysis of Milky Way objects allows for a single velocity profile, but this cannot be done for M31 dwarfs. Andromeda VI and Andromeda XXIII data contain too many MW/M31 contaminants within the line of sight and, therefore, require a multi-Gaussian fit. Four Gaussian peaks are used to fit two MW distributions, a dwarf galaxy distribution, and an M31 distribution. MW contamination can be treated as a single Gaussian distribution, however \citet{Gilbert_2006} finds that it is best fit with two Gaussians. The dwarf member stars are generally distinct from MW populations, as shown in Figures \ref{fig:And6all} and \ref{fig:And23all}. A double-Gaussian approach, however, is consistent with work such as C23 and is used in our analysis. We are then left with velocity dispersion, $\sigma_{\rm{v}}$, and systemic velocity for each field population, $v_{\rm{r}}$. A fit equation is produced to determine a given peak's most probable systemic velocity:

\begin{multline}
\label{eq:P_peak}
P_{\rm{peak}} = \frac{1}{\sqrt{2\pi} \sqrt{(3.2^{2} + \sigma_{\rm{\nu_{peak}}}^{2} + \nu_{\rm{err, i}}^{2})}}\\ 
\times \exp \Bigg(-\frac{1}{2} \Bigg[\frac{\nu_{\rm{peak}} - \nu_{\rm{i}}}{\sqrt{(3.2^{2} + \sigma_{\rm{\nu_{peak}}}^{2} + \nu_{\rm{err, i}}^{2})}}\Bigg]^{2}\Bigg)
\end{multline}
where 3.2 is the inherent uncertainty from  Keck II DEIMOS (see: $\S$ \ref{subsec:DEIMOS}), $\nu_{i}$ is the velocity of a given star, and $\nu_{\rm{err_{i}}}$ is the velocity uncertainty of a given star. This fit is completed for each peak in the data (P$_{\text{gal}}$, P$_{\text{MW1}}$, P$_{\text{MW2}}$, P$_{\text{M31}}$) and then used to construct a logarithmic likelihood equation:

\begin{equation} \label{eq:vel_log_likelihood}
    \log{\mathcal{L}} = \sum_{\rm{i = 1}}^{N} \log \Bigg({-\frac{1}{2}(\alpha P_{\rm{gal}} + \beta P_{\rm{M31}} + \gamma P_{\rm{MW1}} + \delta P_{\rm{MW2}})}\Bigg)
\end{equation}
where $P_{\rm{gal}}$, $P_{\rm{M31}}$, and $P_{\rm{MW}}$ are the peak velocity probabilities of the dwarf galaxy, Andromeda halo contamination, and Milky Way halo contamination, respectively. $\alpha$, $\beta$, $\gamma$, and $\delta$ are constants describing the fraction of stars belonging to each population, normalized such that $\alpha + \beta + \gamma + \delta = 1$. Each component is found via \texttt{emcee} with a routine of 200 walkers for 5000 steps, with a burn-in period of 1550 steps. This routine is similar to that of C21 and C23, where convergence of the walkers occurred prior to the 1550-step cutoff. This cutoff is confirmed via an integrated autocorrelation time, where convergence of 200 walkers is reliably achieved after 1550 steps \citep{Goodman_2010}. The priors used to initialize the runs for Andromeda VI and Andromeda XXIII, as well as the results, can be seen in Table \ref{tab:vel_priors}. Each prior was informed by C13 and C23. $P_{\rm{vel}}$ is then calculated by combining all Gaussian fits for a total velocity probability, $P_{\rm{vel}}$, via the following equation:

\begin{equation}\label{eq:P_vel}
    P_{\rm{vel}} = \frac{P_{\rm{gal}}}{P_{\rm{MW1}} + P_{\rm{MW2}} + P_{\rm{M31}}}
\end{equation}

It is important to note that the resulting values of $v_{\rm{r}}$ and $\sigma_{\rm{v}}$ from Eq. \ref{eq:P_peak} are intermediate and are used only to determine membership likelihood. These likelihoods are then used as weights in our final analysis of the true $v_{\rm{r}}$ and $\sigma_{\rm{v}}$ values. Stars belonging to other stellar populations with similar velocities are more clearly distinguished via $P_{\rm{CMD}}$ calculations and removed. Our final probability is given by combining both these criteria:

\begin{equation} \label{eq:P_total}
    P_{\rm mem} = P_{\rm{CMD}} \times P_{\rm{vel}} 
\end{equation}

\begin{figure}
    \centering
    \includegraphics[width = \columnwidth]{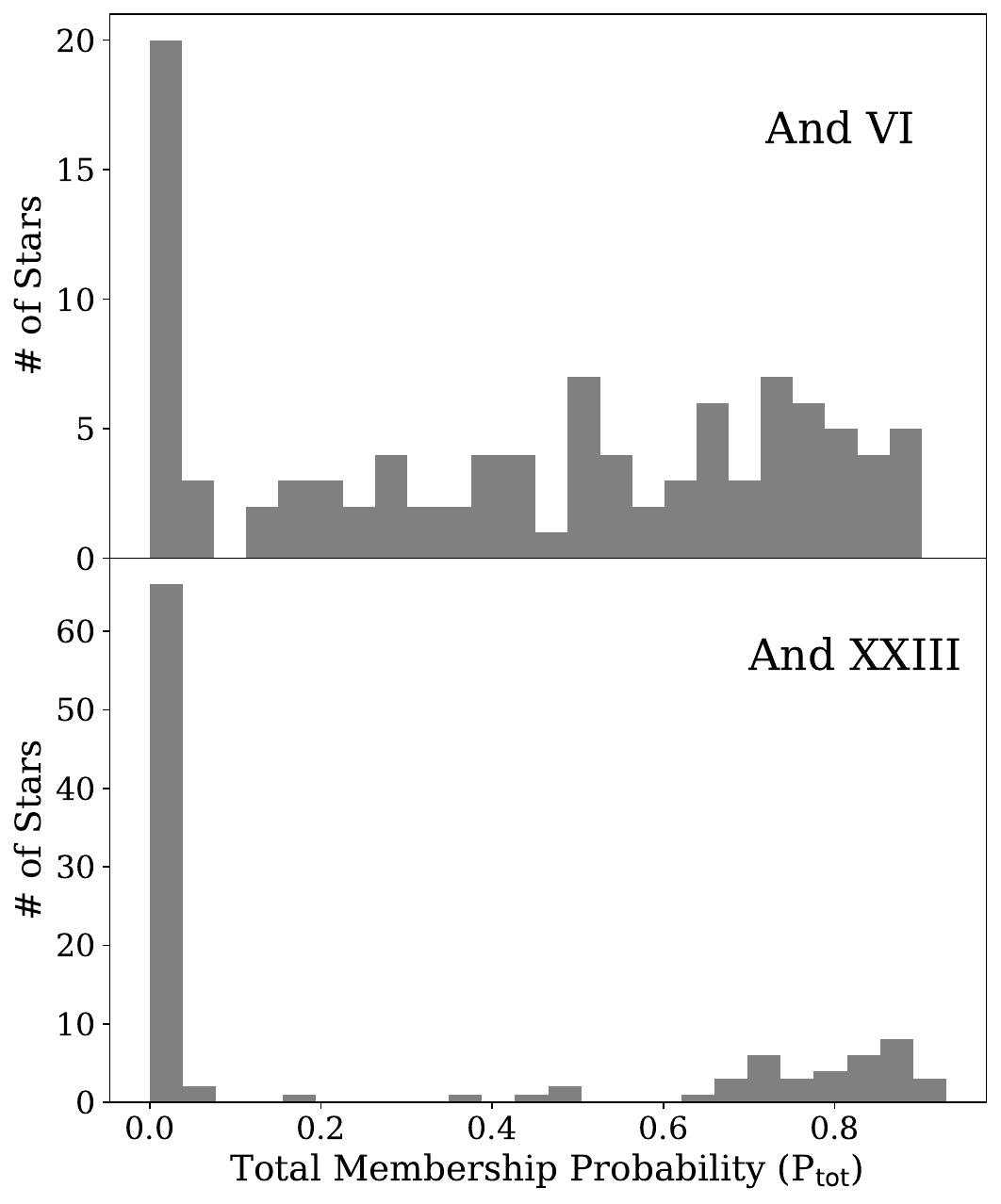}
    \caption{Total probability membership of Andromeda VI (top) and Andromeda XXIII (bottom), where there is a clear distribution bias at $P_{\rm{tot}} < 0.10$. Therefore, only stars with a total membership probability of $P_{\rm{tot}} \geq 0.10$ are considered for our analysis.}
    \label{fig:TotMemHist}
\end{figure}

Stars with a probability of $P_{\rm mem} \geq 0.10$ are considered member stars. A stricter cut would limit potential candidate stars and artificially decrease velocity dispersion estimations, whereas a more lenient cut would artificially inflate the velocity dispersion estimates. This is shown in Figure \ref{fig:TotMemHist}, where no star has a zero probability. However, non-members are noted to have significantly lower membership probabilities than 0.15. With this cut imposed on the data, we identify 75 member stars for Andromeda VI and 39 member stars for Andromeda XXIII.

\begin{table}
\renewcommand{\arraystretch}{1.4}
    \caption{Priors and results of the \texttt{emcee} analysis And VI and XXIII. $\nu_{\rm{r}}$ is the peak velocity and $\sigma_{\rm{\nu}}$ is the velocity dispersion of a given object. These ranges were chosen based off C13 and C23. \textit{Note:} The results presented in this table are only used to determine total membership probabilities. They are not the final systemic velocity and velocity dispersion values for Andromeda VI and Andromeda XXIII.}
    \label{tab:vel_priors}
    \centering
    \begin{tabularx}{\columnwidth}{p{2cm} | Y | Y}
         \textbf{Peak} & $\nu_{\rm{r}}$ (\kms) & $\sigma_{\rm{\nu}}$ (\kms) \\ \hline
         P$_{\rm VI}$ &  -425 $< \nu_{\rm{r}} <$ -300  & 0 $< \sigma_{\rm{\nu}} <$ 20 \\
         P$_{\rm XXIII}$ &  -300 $< \nu_{\rm{r}} <$ -200  & 0 $< \sigma_{\rm{\nu}} <$ 20 \\
         P$_{\rm MW1}$& -100 $< \nu_{\rm{r}} <$ -60 & 0 $< \sigma_{\rm{\nu}} <$ 100 \\
         P$_{\rm MW2}$& -60 $< \nu_{\rm{r}} <$ -20 & 0 $< \sigma_{\rm{\nu}} <$ 100 \\
         P$_{\rm M31}$& -400 $< \nu_{\rm{r}} <$ -130 & 0 $< \sigma_{\rm{\nu}} <$ 200 \\ \hline
         \textbf{Result} & &  \\ \hline
         P$_{\rm VI}$ &  -342.9 $\pm$ 1.8  & 14.0 $\substack{+1.6 \\ -1.5}$ \\
         P$_{\rm XXIII}$ &  -237.9 $\substack{+1.5 \\ -1.4}$  & 6.2 $\pm$ 1.5 
    \end{tabularx}
\end{table}

\subsection{Kinematics of And VI and And XXIII} \label{subsec:kinematics}
We use the results from Table \ref{tab:vel_priors} to determine the true $v_{\rm{r}}$ and $\sigma_{\rm{v}}$ values of both dwarf galaxies. A Gaussian logarithmic likelihood is used to weigh each member stars' probability through a single Gaussian fit:

\begin{equation}\label{eq:VelDispLogLike}
    \log \mathcal{L} = \sum_{i = 1}^{N} \log (P_{\rm mem_{i}}, P_{\rm{gal_i}})
\end{equation}

\begin{figure}
    \centering
    \includegraphics[width = \columnwidth]{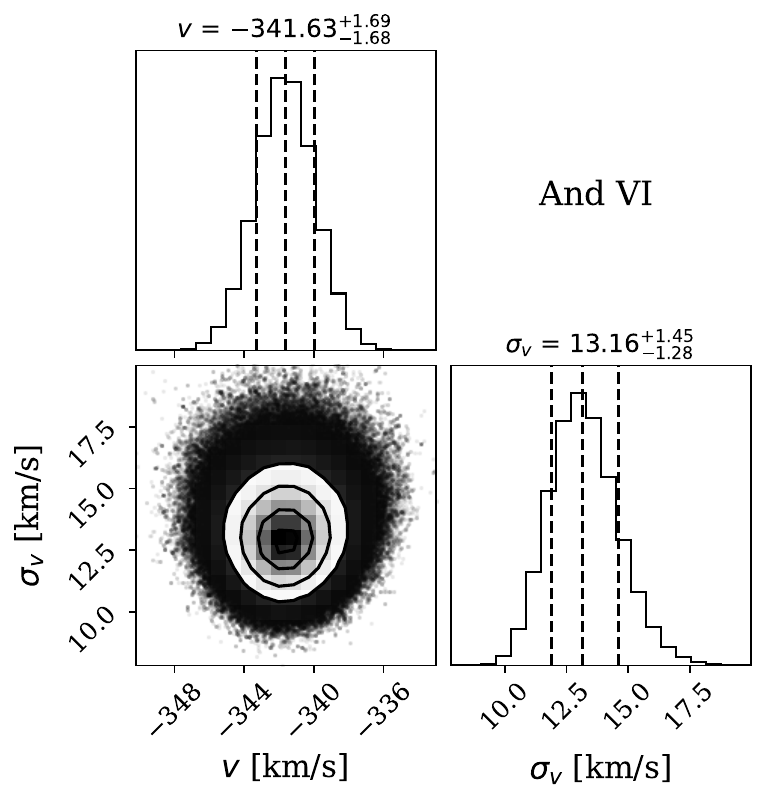}
    \caption{And VI dispersion, as determined via out MCMC analysis routine outlined in $\S$ \ref{subsubsec:velocity}. Both the systemic velocity and veloicty dispersion are shown to converge well at their values of $v$ = (-341.6 $\pm$ 1.7) \kms and $\sigma_{\rm{v}}$ = 13.2 $\substack{+1.4 \\ -1.3}$ \kms, respectively.}
    \label{fig:And6Dispersion}
\end{figure}

\begin{figure}
    \centering
    \includegraphics[width = \columnwidth]{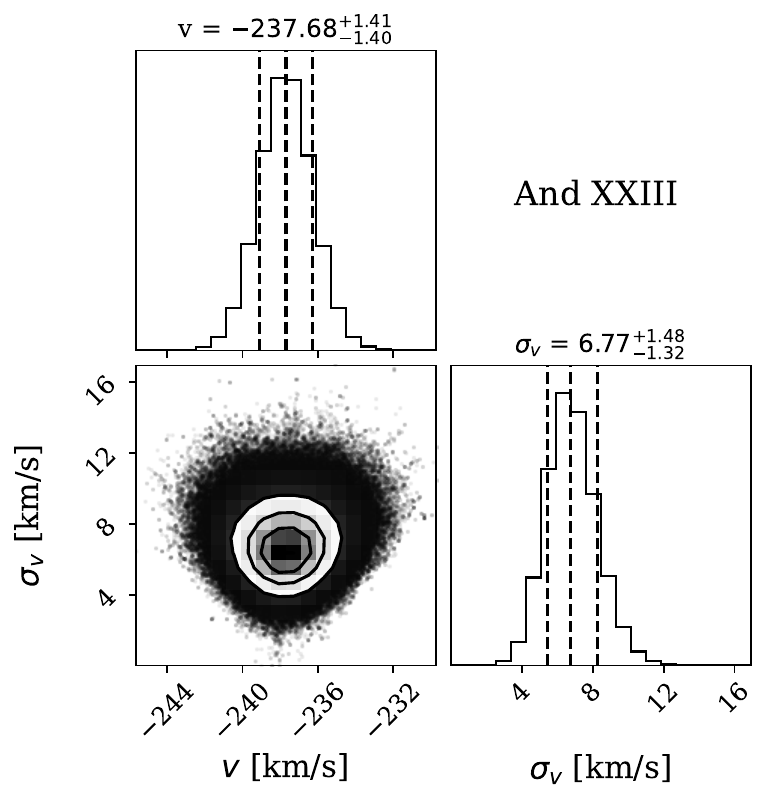}
    \caption{Same as Figure \ref{fig:And6Dispersion} but for Andromeda XXIII. And XXIII's systemic velocity and velocity dispersion are shown to converge well at values of $v$ = (-237.7 $\pm$ 1.4) \kms and $\sigma_{\rm{v}}$ = 6.8 $\substack{+1.5 \\ -1.3}$ \kms, respectively.}
    \label{fig:And23Dispersion}
\end{figure}

This routine is run using \texttt{emcee} with 500 walkers over 5000 steps, with a burn-in period of 2000 steps. This burn-in is also informed by convergence autocorrelation time. The initial guesses for $v_{\rm{r}}$ and $\sigma_{\rm{v}}$ were taken from C13 (noted in Table \ref{tab:vel_data_comparison}), while the same flat priors from Table \ref{tab:vel_priors} were used. This resulted in the posterior distributions shown in Figures \ref{fig:And6Dispersion} and \ref{fig:And23Dispersion}.

Even with a relatively low number of member stars for And VI and And XXIII, we are able to return well-resolved values for systemic velocity and velocity dispersion for both dwarfs, shown in Table \ref{tab:vel_data_comparison}. Velocity values for both And VI and And XXIII fall within 1$\sigma$ of the values found within C13. Some remaining M31 dwarfs are anticipated to have fewer spectroscopic members. Thus, it has yet to be seen how well kinematics are measured via our analysis with fewer member stars than those described here. 


\begin{table}
\renewcommand{\arraystretch}{1.5}
    \centering
    \caption{Systemic velocity and velocity dispersion, as measured in this work, compared to the results from C13.}
    \label{tab:vel_data_comparison}
   \begin{tabularx}{\columnwidth}{p{2cm} | Y | Y}
        \textbf{This work}& $v_{\rm{r}}$ (\kms) & $\sigma_{\rm{v}}$ (\kms) \\ \hline
        And VI & -341.6$\pm$1.7 & 13.2$\substack{+1.4 \\ -1.3}$\\
        AndXXIII & -237.7$\pm$1.4 & 6.78$\substack{+1.5 \\ -1.3}$\\ \hline
        \textbf{C13}& & \\ \hline
        And VI & -339.8$\pm$1.8 & 12.4$\substack{+1.5 \\ -1.3}$\\
        AndXXIII & -237.7$\pm$1.2 & 7.1$\pm$1.0\\ 
    \end{tabularx}
\end{table}

Using our kinematic results, we also recalculate the mass within the half-light radius, $M(r < r_{\rm{h}})$, for each dwarf from their velocity dispersion, using the \citet{Walker_2009} relation. This estimation is reasonably model-independent and thus not affected by $\beta$ anisotropy \citep{Walker_2009, Wolf_2010, Walker_2011}:

\begin{equation} \label{eq:disp_mass}
    M(r < r_{\rm{h}}) = \mu r_{\rm{h}}\sigma^{2}_{\rm{v}}
\end{equation}
where $\mu = 580 ~M_{\rm{\odot}} ~\text{pc}^{-1} ~\text{km}^{-2} ~\text{s}^{2}$.

The enclosed mass of Andromeda VI is determined to be M(r $<$ r$_{\rm{h, VI}}$) = (4.9$\pm$1.5) $\times$ 10$^{7}$ M$_{\odot}$. The mass of Andromeda XXIII is determined to be M(r $<$ r$_{\rm{h, XXIII}}$) = (3.1$\pm$1.9) $\times$ 10$^{7}$ M$_{\odot}$. Figure \ref{fig:rvMenc} displays both dwarfs lying comfortably on the relation created by Eq. \ref{eq:disp_mass}. Using the respective luminosities of And VI and And XXIII $M_{\rm{V, VI}} = -11.6$ and $M_{\rm{V, XXIII}} = -9.8$ \citep{Savino_2022}, we are able to calculate the $[M/L]_{r_{\rm{h}}}$ for each object. Andromeda VI has a luminosity of L$_{r_{\rm{h,VI}}}$ = (18.2 $\pm$0.4) $\times$ 10$^{5}$ L$_{\odot}$ and a mass-to-light ratio of $[M/L]_{r_{\rm{h,VI}}}$ = \MLVI. Andromeda XXIII has a luminosity of L$_{r_{\rm{h,XXIII}}}$ = (3.5$\pm$0.1) $\times$ 10$^{5}$ L$_{\odot}$ and a mass-to-light ratio of $[M/L]_{r_{\rm{h,XXIII}}}$ = \MLXXIII. Our results for Andromeda VI agree with C13, where $[M/L]_{r_{\rm{h,VI}}}$ = 27.5$\substack{+4.2 \\ -3.9}$ M$_{\odot}$/L$_{\odot}$. Our $[M/L]$ ratio for Andromeda XXIII also agrees with C13, where $[M/L]_{r_{\rm{h,XXIII}}}$ = (58.5$\pm$36.2) M$_{\odot}$/L$_{\odot}$. And VI exhibits no sign of gas that contributes to its mass content, as evidenced by both its lack \ce{H}$\alpha$ and \ce{H}II gas \citep{Armandroff_1999} and lack of recent star formation \citep{Weisz_2019, Savino_2025}. And XXIII has  also been observed to be devoid of any gas \citep{Richardson_2011}, with a lack of recent star formation \citep{Weisz_2019, Savino_2025}, further evidencing that its mass content is likely not affected by gas.

\subsection{Mass Profiles of And VI and And XXIII} \label{subsec:massmodels}

In this section, we go beyond previous analysis of both And VI and And XXIII by exploring their mass profiles. The purpose of this process is to better understand the dark matter distribution in both dwarfs, which will allow for comparisons to similar dwarfs in the Milky Way shown in \citet{Read_2016, Read_2017, Read_2018, Read_2019} and \citet{Genina_2020}, as well as to M31 dwarfs such as those studied in C21 and C23.

\subsubsection{Using \texttt{Binulator} and \gravsphere ~for Andromeda VI and Andromeda XXIII} \label{subsec:binulator_gravsphere}




To implement \texttt{Binulator} and \gravsphere ~for both dwarf galaxies, a surface brightness profile (SBP) has to be created using the available photometric data described in $\S$ \ref{subsec:subaru}. Point sources from these data are taken out to 5 $\times ~r_{\rm{h}}$ from the known center of a given dwarf (see Table \ref{tab:dwarf_chars}), with distance membership probability, $P_{\rm{dist}}$, calculated by the following equation:

\begin{equation} \label{eq:P_dist}
    P_{\rm{dist}} = \exp \Bigg(-\frac{1}{2} \bigg(\frac{d^{2}}{r_{\rm{p}}^{2} \eta _{\rm{dist}}^{2}}\bigg)\Bigg)
\end{equation}
where $d$ is the calculated radial distance of each matched star. $\eta _{\rm{dist}}$ is another free scaling parameter, though this time being used to scale the probability to the size of each galaxy. $r_{p}$ is the scale radius of the Plummer profile:

\begin{equation} \label{eq:Plummer_Radius}
    r_{\rm{p}} = \frac{r_{\rm{h}} (1 - \epsilon)}{1 + (\epsilon \cos{\theta_{\rm{i}})}}
\end{equation}
where $r_{\rm{h}}$ is the half-light radius of the dwarf, $\epsilon$ is the ellipticity of the dwarf, and $\theta_{\rm{i}}$ is the matched star's inclination with respect to the dwarf's major axis \citep{Plummer_1911}. The Plummer profile modifies the half-light radius depending on the ellipticity of the dwarf. Parameters used to calculate $P_{\rm{dist}}$ are shown in Table \ref{tab:P_dist_values}.

Constraints of $\eta_{\rm{dist}} = 5$ and $P_{\rm{dist}} > 0.01$ are placed on the SBP data to ensure a robust set of members and background, mirroring that of the process in C23. The velocity of each member is determined using the available spectroscopic data, described in $\S$ \ref{subsec:DEIMOS}. The number of effective dynamical tracers for both systems is calculated as follows:

\begin{equation} \label{eq:tracer_number}
    N_{\text{eff}} = \sum^{N_{\rm{mem}}}_{i = 0} P_{\rm{mem, i}}
\end{equation}
where the total $N$ value is the sum of probabilities for each star. And VI is found to have a total number of 5398 photometric tracers, split into 70 bins of 77 stars, and 39 velocity tracers, split into 6 bins of 6 stars. And XXIII is found to have 1490 photometric tracers, split into 49 bins of 30 stars, and 28 velocity tracers, split into 7 bins of 4 stars. The SBPs for And VI and And XXIII can be seen in Figures \ref{fig:And6SBP} and \ref{fig:And23SBP} in Appendix \ref{subsec:append_plots}. It should be noted that a distance limit of 900 pc was applied to the photometric data of Andromeda VI prior to the creation of the SBP. This was done to prevent background stars, which appeared prominently above the distance limit, from being fit. A distance limit of 1800 pc was applied to the photometric data of And XXIII for a similar reason.

\gravsphere ~fits the SBP, as well as the radial velocity profile, created through the \texttt{Binulator} binning process. While accounted for when fitting the SBP, final DM density estimations from \gravsphere ~do not change based on solar distance measurements to the dwarf. However, it should be noted that using measured uncertainties may help marginalize better over distance. The \texttt{CoreNFWTides} parameters used to fit each galaxy can be seen in Table \ref{tab:CoreNFW_priors}. Stellar mass is obtained via the updated luminosity values shown in \citet{Savino_2022}, using the stellar mass-to-light ratio assumption of 2 for old stellar populations \citep{Simon_2019}. We obtain a stellar mass for Andromeda VI of M$_{\rm{*,VI}}$ = $(7.3 \pm 0.2) \times 10^{6}$ M$_{\odot}$ and a stellar mass for Andromeda XXIII of M$_{\rm{*,XXIII}}$ = $(1.4 \pm 0.04) \times 10^{6}$ M$_{\odot}$. Full diagnostic plots for Andromeda VI and Andromeda XXIII can be seen in Appendix \ref{subsec:append_plots} as Figures \ref{fig:And6VelArray} and \ref{fig:And23VelArray}, respectively.

\begin{table}
\renewcommand{\arraystretch}{1.5}
    \caption{$P_{\rm{dist}}$ parameters for And VI and And XXIII, taken from C13 and \citet{Martin_2016}, respectively.}
    \label{tab:P_dist_values}
    \centering
    \begin{tabularx}{\columnwidth}{X | Y | Y | Y}
    \textbf{Galaxy} & $r_{\rm{h}}$ (arcmin) & $\epsilon$ & $\theta _{\rm{i}}$ ($^{\circ}$) \\ \hline
    And VI & 1.9 $\pm$ 0.08 & 0.39 $\pm$ 0.03 & 164 $\pm$ 3 \\
    And XXIII & 5.4 $\pm$ 0.4 & 0.41$\substack{+0.05 \\ -0.06}$ & 138 $\pm$ 5 \\
    \end{tabularx}
\end{table}

\begin{table}
\renewcommand{\arraystretch}{1.4}
    \centering
    \caption{Priors used for the  \texttt{CoreNFWTides} models of Andromeda VI and Andromeda XXIII in \texttt{GRAVSPHERE}.$M_{\rm{200}}$ is the enclosed mass at the virial radius. $c_{\rm{200}}$ is the dimensionless concentration parameter. $r_{\rm{c}}$ represents the size of a constant density core, and $r_{\rm{t}}$ represents the outer `tidal radius' at which density fall-off steepens at a rate of $\rho \propto r^{-\delta}$. $n$ sets the shape of the dark matter density profile, with $n = 0$ corresponding to a central constant density core and $n = 1$ representing a cusp of $\rho \propto r^{-1}$. $\tilde\beta_{\rm{0}}$ is the central symmetrized velocity anisotropy profile. $\tilde\beta_{\rm{\infty}}$ is the outer symmetrized velocity anisotropy profile ($\tilde\beta = -1$ represents a fully tangential distribution, $\tilde\beta = 1$ represents a fully radial distribution. Our priors on $\tilde\beta_{\rm{0}}$ and $\tilde\beta_{\rm{\infty}}$ are broad while being motivated by the latest cosmological simulations in the standard cosmology. These find that the stellar velocity anisotropy should be relatively isotropic in the center and radial at larger radii (e.g. \citet{Orkney_2023}). $r_{\rm{0}}$ is a transition radius, with $q$ controlling the transition sharpness. See \citet{Read_2018} and \citet{Collins_2021} for further details of \texttt{CoreNFWTides} parameters.} 
    \label{tab:CoreNFW_priors}
    \resizebox{\columnwidth}{!}{%
    \begin{tabularx}{\columnwidth}{p{1.5cm} | Y | Y}
       \textbf{Priors} & Andromeda VI & Andromeda XXIII \\ \hline
       $\log_{10} (\frac{M_{\rm{200}}}{M_{\odot}})$ & $9.25 {<} \log_{10} (\frac{M_{\rm{200}}}{M_{\odot}}) {<} 10.5$ & $8.5 {<} \log_{10} (\frac{M_{\rm{200}}}{M_{\odot}}) {<} 10.0$ \\
       
       $c_{200}$ & $7 < c_{200} < 53$ & $7 < c_{200} < 53$ \\
       
       $\log_{10}(\frac{r_{\rm{c}}}{\text{kpc}})$ & $0.01 < \log_{10}(\frac{r_{\rm{c}}}{\text{kpc}}) < 10$ & $0.01 < \log_{10}(\frac{r_{\rm{c}}}{\text{kpc}}) < 10$ \\

       $\log_{10}(\frac{r_{\rm{t}}}{\text{kpc}})$ & $1 < \log_{10}(\frac{r_{\rm{t}}}{\text{kpc}}) < 20$ & $1 < \log_{10}(\frac{r_{\rm{t}}}{\text{kpc}}) < 20$ \\

       $\delta$ & $3.01 < \delta < 4$ & $3.01 < \delta < 4$ \\

       $n$ & $0 < n < 1$ & $0 < n < 1$ \\ \hline \hline

       $\tilde{\beta}_{0}$ & $-0.01 < \tilde{\beta}_{0} < 0.01$ & $-0.01 < \tilde{\beta}_{0} < 0.01$ \\
       
       $\tilde{\beta}_{\infty}$ & $-0.1 < \tilde{\beta}_{\infty} < 1$ & $-0.1 < \tilde{\beta}_{\infty} < 1$ \\
       
       $c_{200}$ & $7 < c_{200} < 53$  & $7 < c_{200} < 53$ \\
       
       $\log_{10}(\frac{r_{\rm{0}}}{\text{kpc}})$ & $-2 < \log_{10}(\frac{r_{\rm{0}}}{\text{kpc}}) < 0$ & $-2 < \log_{10}(\frac{r_{\rm{0}}}{\text{kpc}}) < 0$\\

       $q$ & $1 < q < 3$ & $1 < q < 3$ \\
    \end{tabularx}}
\end{table}

\subsection{Dark Matter Densities} \label{subsec:dm_densities}
\gravsphere ~outputs the estimated dark matter densities of a given system. These are visualized in Figures \ref{fig:And6_SFR_Den} and \ref{fig:And23_SFR_Den}. Each image shows the 1$\sigma$ uncertainty in dark gray and 2$\sigma$ uncertainty in light gray. The half-light radius is denoted by the solid blue vertical line. A distance of 150 pc from the center of the dwarf, where dark matter density is measured, is denoted by the pink dashed line and is a DM measurement distance consistent with \citet{Read_2018}, \citet{Genina_2020}, C21, and C23.

\begin{figure}
    \centering
    \includegraphics[width=\columnwidth]{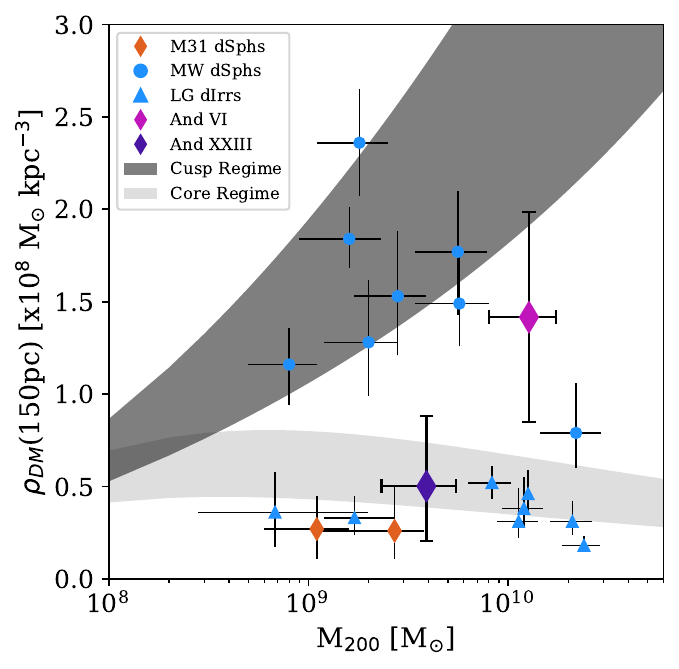}
    \caption{Central dark matter density as a function of pre-infall halo mass, $M_{\rm{200}}$. The dark gray region represents a fully cusped profile, while the light gray region represents a fully cored profile (\texttt{coreNFW} from \citet{Read_2016}). Each band has a width corresponding to a 1$\sigma$ scatter in the DM halo concentration \citep{Dutton_2014}. Milky Way dwarf spheroidal galaxies are shown as light blue circles, while Local Group dwarf irregulars are shown as light blue triangles. Andromeda dwarfs spheroidals are shown as diamonds, with previously-mass modeled dwarfs being shown in orange. Andromeda VI is represented as a magenta diamond, while Andromeda XXIII is represented as a dark purple diamond. The error bars for each point are 1$\sigma$ uncertainties. Data for other M31 dwarfs were obtained from C21 and C23. Data for MW and Local Group dwarfs were obtained from \citet{Read_2019}.}
    \label{fig:M200_DM_comp}
\end{figure}

Analysis of Andromeda VI returns a central dark matter density at 150 pc of \DMVI. Analysis of Andromeda XXIII returns a dark matter density at 150 pc of \DMXXIII. Andromeda VI falls within the higher, "cuspy" regime of dwarf galaxies, while Andromeda XXIII appears to reside in the lower, "cored" regime, similar to that of And XXI and And XXV. Comparisons to more Local Group systems can be seen in Figure \ref{fig:M200_DM_comp}, while causes for these profiles are further explored in $\S$\ref{subsec:starformation}.

\subsection{Star Formation History} \label{subsec:starformation}


The Star Formation History (SFH) can reveal important information about changes to a dwarf galaxy's dark matter density profile. Extended star formation can potentially lead to more significant changes, while dark matter density appears to remain relatively unaffected in systems that have had their star formation shut down (quenched) early in their lifetime \citep{Moster_2010, Kirby_2011, Brown_2014, Read_2016, Katz_2017, Read_2017, Kravtsov_2018, Read_2018, Weisz_2019, Charles_2022, Savino_2023, Muni_2025}. It can be difficult to link stellar and halo masses in a dwarf if it has not continuously formed stars. A quenched system's stellar mass may be "frozen in," creating scatter in the M$_{*}$-M$_{\rm{200}}$ relation for isolated dwarfs  \citep{Read_2017, GarrisonKimmel_2017, Nadler_2020, Manwadkar_2022, Danieli_2023, OLeary_2023, Ahvazi_2024, Kim_2024}. This scatter can complicate attempts to understand dark matter density based only on M$_{*}$ measurements, as described in \citet{ReadErkal_2019}. Looking at the entire SFH of a dwarf can provide a more robust picture of its mass relations and dark matter density.

\begin{figure}
    \centering
    \includegraphics[width = \columnwidth]{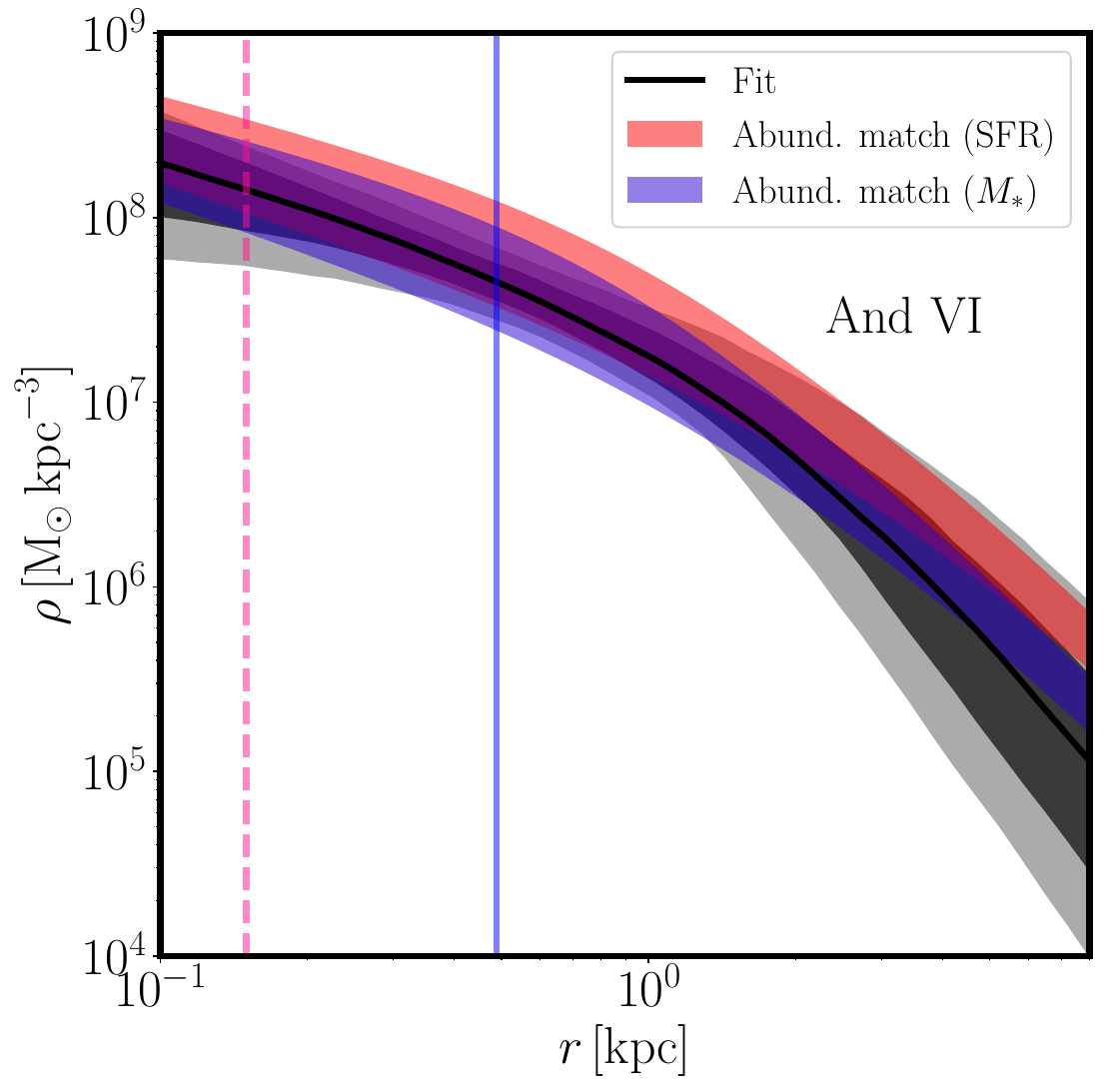}
    \caption{Andromeda VI Dark matter density profiles determined by \gravsphere ~(solid black line), with 1 and 2$\sigma$ being denoted by the dark gray and light gray shaded regions, respectively. The pink shaded region is the density profile created via $\langle{\text{SFR}}\rangle$ abundance matching. The blue shaded region is the density profile created via $M_{\rm{*}}$ abundance matching. The blue vertical line is the measured $r_{\rm{h}}$ of And VI. The pink dashed line is at 150 pc, or the distance at which DM density is measured.}
    \label{fig:And6_SFR_Den}
\end{figure}

\begin{figure}
    \centering
    \includegraphics[width = \columnwidth]{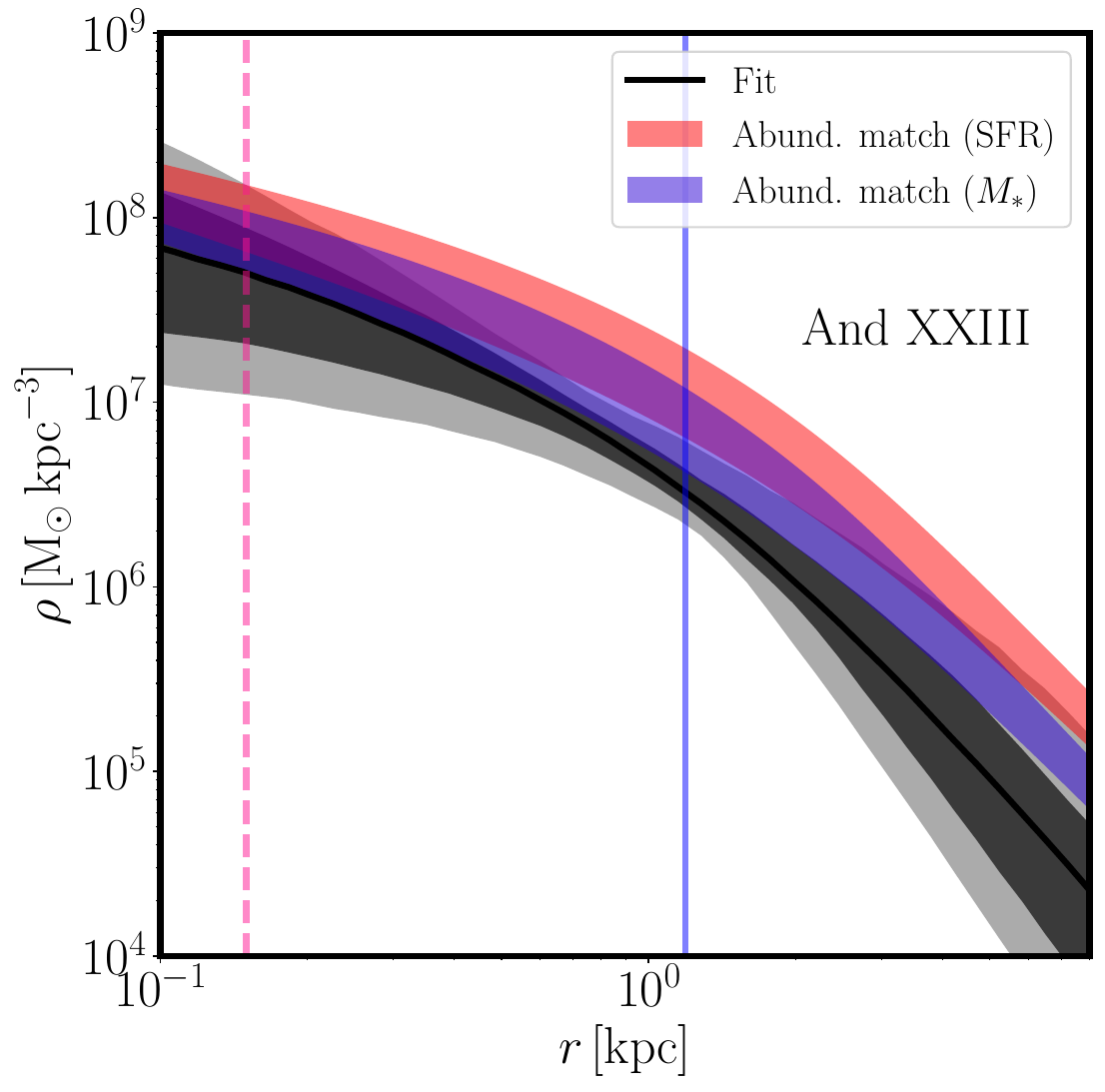}
    \caption{Same as Figure \ref{fig:And6_SFR_Den} but for Andromeda XXIII.}
    \label{fig:And23_SFR_Den}
\end{figure}

The star formation histories of And VI and And XXIII indicate that many of their stars were formed early on in their lifetime. New, deeper data presented by \citet{Savino_2025} reveals that And VI formed 90\% of its stellar population $\sim$6.8 Gyr ago, while And XXIII formed 90\% of its stars $\sim$7.3 Gyr ago. Star formation therefore occurred in both dwarfs for about 50\% of the lifetime of the Universe. However, both quenched shortly afterwards and have shown no signs of significant star formation since. The measured DM density of And VI can be explained via its SFH, as seen in Figure \ref{fig:And6_SFR_Den}. Conversely, while And XXIII's SFH can explain its lower DM density to a degree, Figure \ref{fig:And23_SFR_Den} shows that it cannot be explained through SFH alone. Therefore, this significant deviation from the expected density is most likely not due to internal effects, such as DM heating due to supernovae.

The process detailed in \citet{ReadErkal_2019} uses SFR matching to determine the DM density of central and isolated satellite galaxies, assuming minimal disruptions to the system. It is shown that the mean SFR, $\langle{\text{SFR}}\rangle$, monotonically rises with the virial mass, $M_{\rm{200}}$. For satellite galaxies, such as And VI and And XXIII, this produces much less scatter than an $M_{*}-M_{\rm{200}}$ relation. The blue bands Figures \ref{fig:And6_SFR_Den} and \ref{fig:And23_SFR_Den} represent the expected DM density using an abundance-matched $M_{\rm{200}}$ value based on $M_{\rm{*}}$. The red bands take into account the `star formation time,' described in \citet{ReadErkal_2019} as $t_{\rm{SF}} = M_{\rm{*}}/\langle{\text{SFR}}\rangle$, in an effort to determine core formation based on $\langle{\text{SFR}}\rangle$. Coring of the DM density is then calculated as $n = \tanh(0.04t_{\rm{SF}}/t_{\rm{dyn}})$, where $t_{\rm{dyn}}$ is the dynamical time at a scale radius, $r_{\rm{s}}$ \citep{Read_2016}. The latter method does not account for tidal mass loss, though neither dwarf detailed in this paper has had its respective orbit constrained. While both currently appear to be at far enough distance from M31 to have little interaction, they could have had a closer pass at some previous time.

And VI and And XXIII each formed 90\% of their stars nearly 7 Gyr ago, with no clear bursts of star formation since. This aligns with other M31 dSphs, most of which formed 90\% of their stars $>5$ Gyr ago, with the exception of Andromeda XVII, XIV, and XII, which formed 90\% of their stars $<5$ Gyr ago \citep{Weisz_2019}. Neither And VI nor And XXIII show indicators of DM heating or cusp-core transformations due to bursty star formation, which is visualized when comparing the dark matter density plots determined by \gravsphere, $M_{\rm{*}}$ abundance matching, and $\langle{\text{SFR}}\rangle$ abundance matching. As seen in Figure \ref{fig:And6_SFR_Den}, each of the three curves produced for And VI are in agreement with one another, particularly at $\leq$150 pc, where all profiles lie within 1$\sigma$ of each other at the $r_{\rm{h}}$. This indicates that And VI has had little disturbance in its mass and has not likely recently interacted with M31. The abundance matched profiles of Andromeda XXIII agreed less with each other, where the current DM density agrees with only $M_{\rm{*}}$ matching within 2$\sigma$, as seen in Figure \ref{fig:And23_SFR_Den}. The curve produced by $\langle{\text{SFR}}\rangle$ matching appears above 2$\sigma$ of the current DM profile but agrees within 1$\sigma$ of $M_{\rm{*}}$ matching. This can be seen in Figure \ref{fig:And6_SFR_Den}. This tensions suggests that And XXIII has had its DM density lowered by some other method. This disparity between these profiles is further discussed in $\S$ \ref{sec:discussion}.

\section{Discussion} \label{sec:discussion}
Andromeda VI has revealed itself to be a typical, compact dwarf galaxy within the M31 system. With no recent star formation and a distance to its host of 281.6$\substack{+8.6 \\ -7.1}$ kpc, the dSph does not seem to have a turbulent history \citep{Weisz_2019, Savino_2022}. The dark matter density aligns with those expected from both  $\langle{\text{SFR}}\rangle$ and M$_{*}$ abundance matching \citep{Read_2016, ReadErkal_2019}. Figure \ref{fig:M200_DM_comp} shows that And VI lies, within uncertainty, nearer the cusped regime, alongside MW systems like Sculptor and Leo I.

\begin{figure}
    \centering
    \includegraphics[width = \columnwidth]{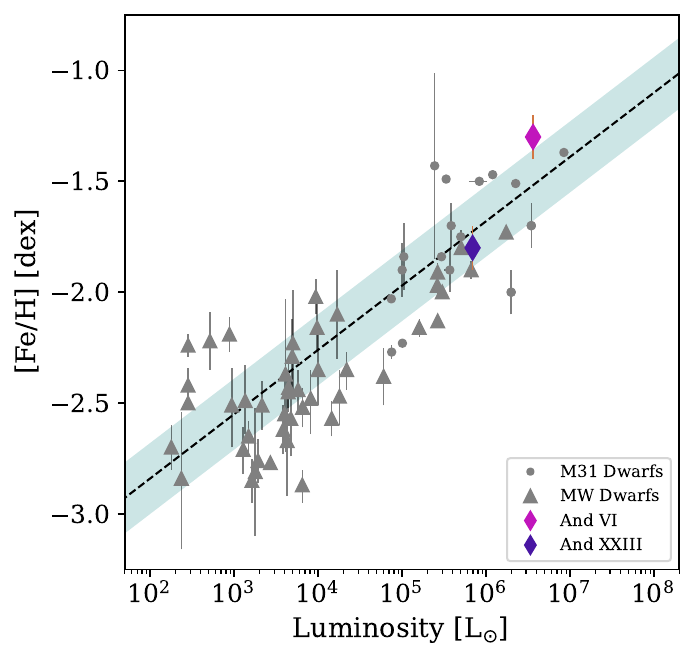}
    \caption{The luminosity-metallicity relation for Local Group dwarf galaxies. Milky Way satellites are denoted by gray triangles, while M31 satellites are denoted by gray circles. The dashed black line represents the relation found by \citet{Kirby_2013}, with the teal shading representing the fit's 1$\sigma$ scatter. Both Andromeda VI and Andromeda XXIII, shown respectively as a magenta diamond and dark purple diamond, lie comfortably within the relation. The data shown in gray are from \citet{Pace_2024}.}
    \label{fig:L_FeH_comp}
\end{figure}

Conversely, Andromeda XXIII presents a lower dark matter density profile than what abundance matching and dark matter heating may suggest. Despite this, a low DM density could very well be explained by tidal interactions with M31. The lowering of the density would be difficult to reconcile with any starbursts or extended star formation \citep{Weisz_2019}. While \citet{Navarro_1996a} initially theorized that a single, explosive mass-loss event could sufficiently explain dark matter cores, works such as \citet{Read_2005} and \citet{Pontzen_2012} instead show that a single starburst is unlikely to contribute enough energy to significantly lower the DM density at the mass scale of And XXIII. `More frequent and energetic' bursts, which are not seen in the SFH of And XXIII, would be necessary to explain coring. This is further explained in \citet{Muni_2025}, where it is shown that `sustained cycles of star formation' are required to cause long-term gravitational potential fluctuations and cusp softening. It is possible that, at a distance to its host of 128.1$\substack{+10 \\ -4.9}$ kpc \citep{Martin_2016, Savino_2022}, And XXIII could have had a previous M31 infall where its mass was stripped. The dwarf is also extended, with a half-light radius of \rhXXIII, which may be a result of interactions with its host. The metallicity of And XXIII is in agreement with the luminosity-metallicity relation of dwarfs found in \citet{Kirby_2013}, as shown in Figure \ref{fig:L_FeH_comp}.

While strong tides can remove stars, causing systems to deviate off the mass-metallicity relation \citep[i.e. Sagittarius dSph]{Ibata_1994}, this does not seem to be the case for And XXIII. Tides that are not strong enough to strip stars may only strip dark matter from the dwarf, allowing for the retention of metallicity and having little effect on the system's luminosity. Thus, our analysis does not preclude the possibility of tides strong enough to lower the dwarf's DM density, like in Crater II and Antlia II. Unfortunately, the dwarf does not seem to resemble many Milky Way dwarfs in any of the comparisons used in our analysis, leaving it difficult to draw conclusions about its structure and DM content. It could be likely that this dwarf has had similar histories as And XXI and And XXV, since all fall within a similar DM density regime \citep{Collins_2021, Charles_2022}. As described in C23 for And XXV, it is important to consider any tidal substructure in dwarfs to more conclusively understand their history. Without proper motion measurements or deep imaging of many M31 dwarfs, however, analysis regarding their tidal interaction histories is not straightforward.

Cold Dark Matter is able to explain the central density of Andromeda VI well, but alternative dark matter models may better explain the lower central density of Andromeda XXIII. Cores can naturally arise within the Self-Interacting Dark Matter paradigm via gravothermal core-collapse \citep{Read_2018, Correa_2021}. Conversely, fermionic warm dark matter could lead to a naturally formed DM core as a consequence of self-gravitating DM \citep{Dalcanton_2001, DiPaolo_2018, Alvey_2021, Arguelles_2023}. This pairs with minimal velocity dispersion, which could align with the low $\sigma_{\rm{v}}$ value of And XXIII. Warm dark matter models have also shown weaker cusps, leading to a lower central density, though the results strongly depend on the particle mass used \citep{Avila_Reese_2001, Maccio_2012, Shao_2013, Lovell_2014, Schneider_2017}. Bosonic dark matter, on the other hand, is theorized to create `false,' solitonic cores, where axions are the cause of cuspy dwarfs being fit well by cored NFW profiles \citep{GonzalezMorales_2017, Hiyashi_2021, Zimmerman_2024}. While mechanisms within CDM may be able to explain the lower-than-expected density of And XXIII, other dark matter models may be able to do so better.

A dwarf galaxy void of dark matter could appear to have a lower central mass and density. Therefore, Modified Newtonian Dynamics (MOND) \citep{McGaugh_2013}, which uses dark matter-free calculations to explain the kinematics of Andromeda dwarf galaxies, must also be considered. The velocity dispersion for Andromeda VI via MOND is estimated to be $\sigma_{\rm{VI,MOND}} = 9.4 \substack{+3.2 \\ -2.4}$ \kms, in agreement with our presented value within 1$\sigma$. The same can be said for And XXIII, where $\sigma_{\rm{XXIII,MOND}} = 6.4 \substack{+1.2 \\ -1.0}$ \kms. A second measurement is also provided, accounting for And XXIII being embedded in an external field in the context of the `external-field effect' (EFE). The velocity dispersion for And XXIII in an EFE context is presented as $\sigma_{\rm{EFE}} = 4.4 \substack{+1.8 \\ -1.3}$ \kms, lower than our measurement but in agreement within 1$\sigma$. MOND therefore agrees with our results.

\subsection{M31 Dwarf System}
The low DM profile of And XXIII makes it the third mass-modeled M31 dwarf spheroidals to present a lower-than-expected DM density. Along with And XXI and And XXV \citep{Collins_2021, Charles_2022}, low-DM dwarfs make up 10\% of the 35 observed satellites around Andromeda. It has yet to be seen, via mass modeling, how many of the remaining 31 Andromeda dSphs also fall under a low-density designation. If a larger amount emerges, then the question of why the Milky Way dwarf population does not resemble Andromeda dwarf population must be investigated as well.

As previously mentioned, without proper motion (PM) measurements, it is difficult to fully understand the orbits of M31 systems. While more M31 satellite PMs are gradually being constrained, these studies can take decades to complete, with only three systems (NGC 147, NGC 185, and Andromeda III) being measured so far \citep{Sohn_2020, Pawlowski_2021}. PM measurements of Andromeda III were only recently estimated using 22 years of nonconsecutive \textit{HST} data, spanning from 1999 to 2021 \citep{Casetti_2024}. Currently, constraints have not yet been placed on any of the mass-modeled dwarf galaxies. Orbit tracing can grant a deeper understanding of effects, such as tides, over the lifetime of a dwarf galaxy. Without these measurements, it is difficult to determine a direct cause of these lower densities.

Further analysis will determine how common these low-density dwarfs are in the M31 satellite population. Planned additions to \gravsphere ~will implement more realistic tidal stripping parameters to further consider its effect on dwarf galaxies. For example, progressive central density lowering due to tides will be considered. The current version of \gravsphere ~only fits for an assumed truncated DM density profile. An updated version will be paired with updated star formation histories using deeper data than those in \citet{Weisz_2019}. These improvements will bolster the analysis discussed in $\S$ \ref{subsec:dm_densities}, where the average SFR of a dwarf can reveal its undistrubed DM density profile \citep{ReadErkal_2019}. With these improvements to data and software, it will be possible to understand the profiles of LG dSphs with more precision than previous work. 



\section{Conclusions} \label{sec:conclusions}
In this paper, we presented updated kinematic analyses of Andromeda VI and Andromeda XXIII. Both are brighter M31 dwarfs with M/L ratios that fall within the expected range for their measured size. Our data set included 75 member stars for And VI and 39 member stars for And XXIII, both of which were sufficient to measure the velocity profiles and dark matter densities of both dwarfs. The density profiles for both systems were estimated using the dynamical mass modeling package of \texttt{Binulator}+\gravsphere. Our key findings were as follows:

\begin{itemize}
    \item {We measured a systemic velocity of $v = (-341.6 \pm 1.7$) \kms and a velocity dispersion of $\sigma_{\rm{v}} = 13.2 \substack{+1.4 \\ - 1.3}$ \kms for Andromeda VI. We also measure a systemic velocity of $v = (-237.7 \pm 1.4$) \kms and a velocity dispersion of $\sigma_{\rm{v}} = 6.8 \substack{+1.5\\ - 1.3}$ \kms for Andromeda XXIII. These values are consistent with those found in C13.}

    \item{We assumed that both dwarfs are in dynamical equilibrium and used this to determine the mass contained within their half-light radius. This presented a value of $M(r < r_{\text{h}}) = (4.9 \pm 1.5) \times 10^{7}$ and $M(r < r_{\text{h}}) = (3.1 \pm 1.9) \times 10^{7}$ M$_{\odot}$ for And VI and And XXIII, respectively. Andromeda VI has a [M/L]$_{r_{\text{h}}}$ ratio of \MLVI, and Andromeda XXIII has a [M/L]$_{r_{\text{h}}}$ ratio of \MLXXIII. Both And VI and And XXIII exhibited ratios indicative of DM dominance, even within uncertainties.}

    \item{The dynamical mass modeling tool \texttt{Binulator +} \gravsphere ~revealed that Andromeda VI has an expected dark matter density for its [M/L] ratio and size, compared to the Standard Cosmology, at \DMVI. Conversely, mass modeling has showed that Andromeda XXIII presents a lower DM density, at \DMXXIII. This lower density is similar to that of Andromeda XXI and Andromeda XXV. And XXIII's lower density appears to be from external forces rather than internal starbursts, with a quenched SFH.}

    \item{Mass modeling of the remaining dwarf galaxies will be presented in future companion works, with the goal of comparing M31 satellites to those of the Milky Way. In doing so, further effects may be explored to explain the lower densities of these dwarfs, should more appear.}
\end{itemize}


\section{Acknowledgments} \label{sec:acknowledgments}

CSP acknowledges support from STFC studentship grant ST/X508810/1. MLMC acknowledges support from STFC grants ST/Y002857/1 and ST/Y002865/1.

CSP acknowledges Stacy Y. Kim, a Nashman Fellow at Carnegie Science in Pasadena, California, for her help in understanding the role of tidal forces in \gravsphere ~estimations.


\section{Data Availability} \label{sec:dataavailability}
PAndAS and Subaru Suprime-Cam photometric data are available via the Canadian Astronomy Data Centre (CADC) archive. Raw spectra obtained with DEIMOS are available via the Keck archive. Fully reduced, one-dimensional spectra and photometry will be made available upon reasonable request to the lead author. Electronic tables with reduced properties (coordinates, magnitudes, velocities, and membership probability) for all stars and identified member stars will be provided on the journal website. The updated \gravsphere ~\texttt{v1.5} code, with the appropriate \texttt{Binulator} binning method, is available at \url{https://github.com/justinread/gravsphere}.

\bibliographystyle{mnras}
\bibliography{BibFile} 




\appendix \label{sec:appendix}

\section{Supplementary and Diagnostic Plots}\label{subsec:append_plots}

\begin{figure}
    \centering
    \includegraphics[width=\columnwidth]{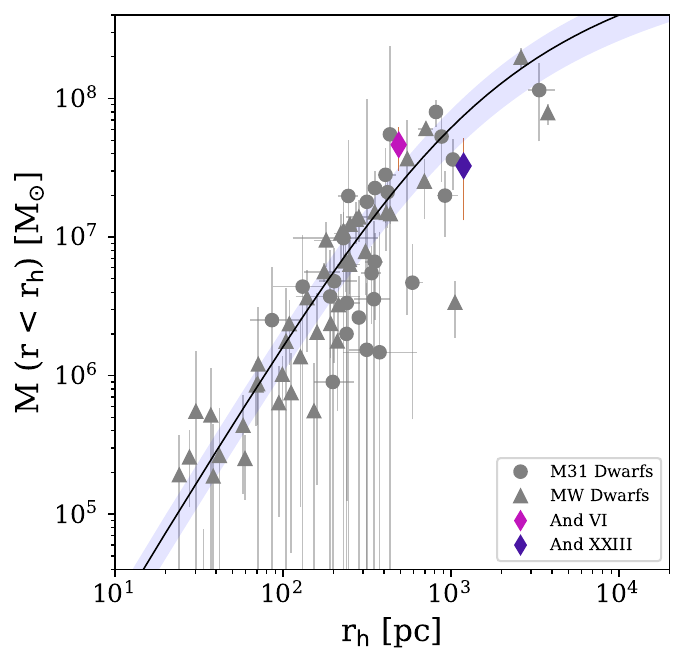}
        \caption{Radius versus enclosed mass of Local Group dwarf galaxies. Milky Way dwarfs are denoted by gray triangles, while M31 dwarfs are denoted by gray circles. Andromeda VI and Andromeda XXIII are represented by a magenta diamond and a dark purple diamond, respectively. The black line represents the best-fit model, from C13, with the blue shading representing 1$\sigma$ uncertainty. Note that some uncertainties are too small to be seen but are represented by the size of the markers. The data shown in gray are from \citet{Read_2019}, \citet{Savino_2022}, and \citet{Pace_2024}.}
    \label{fig:rvMenc}
\end{figure}

This section contains plots that further explain relations and fits described within the text. Figure \ref{fig:rvMenc} displays the $r_{\rm{h}} - M(r < r_{\rm{h}})$ relation discussed in $\S$ \ref{subsec:kinematics}. Comparisons of Andromeda VI and Andromeda XXIII to Local Group dwarfs show that they fall within an expected enclosed mass range for their respective sizes. The black line in the figure represents the best-fit model, from C13, with the blue shading representing 1$\sigma$ uncertainty. Both dwarfs described in this paper lie on the model curve, within uncertainties, indicating that their enclosed mass is appropriate for their size.

\begin{figure}
    \centering
    \includegraphics[width = \columnwidth]{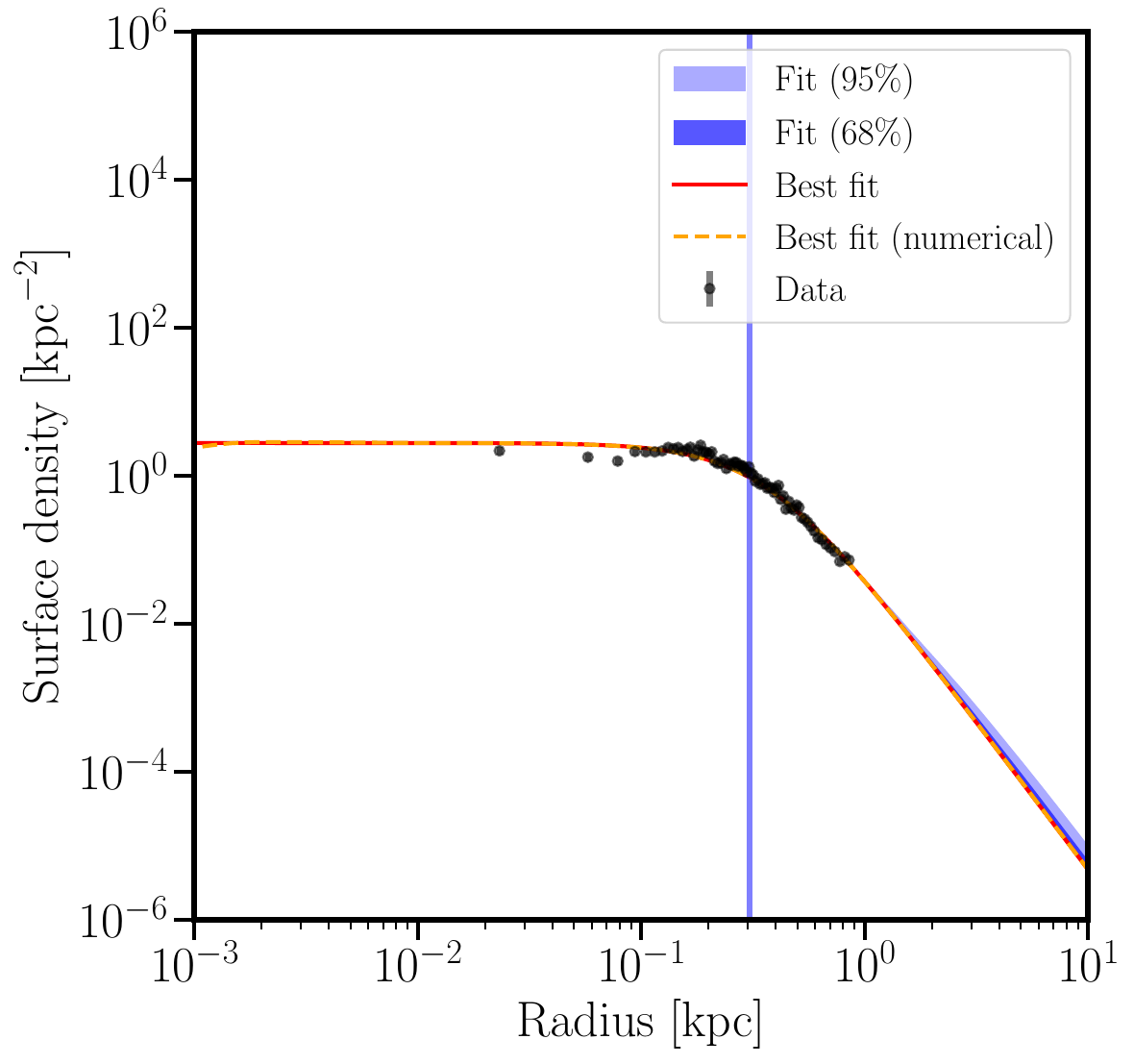}
    \caption{Surface brightness profile of Andromeda VI, as fit by \texttt{Binulator}. The black circles are binned data, with uncertainties being represented by the size of each marker. The vertical blue line is the estimated half-light radius as determined by \texttt{Binulator}. An artificial cut of photometric data was performed at 900 pc to prevent the fitting of background stars.}
    \label{fig:And6SBP}
\end{figure}

\begin{figure}
    \centering
    \includegraphics[width = \columnwidth]{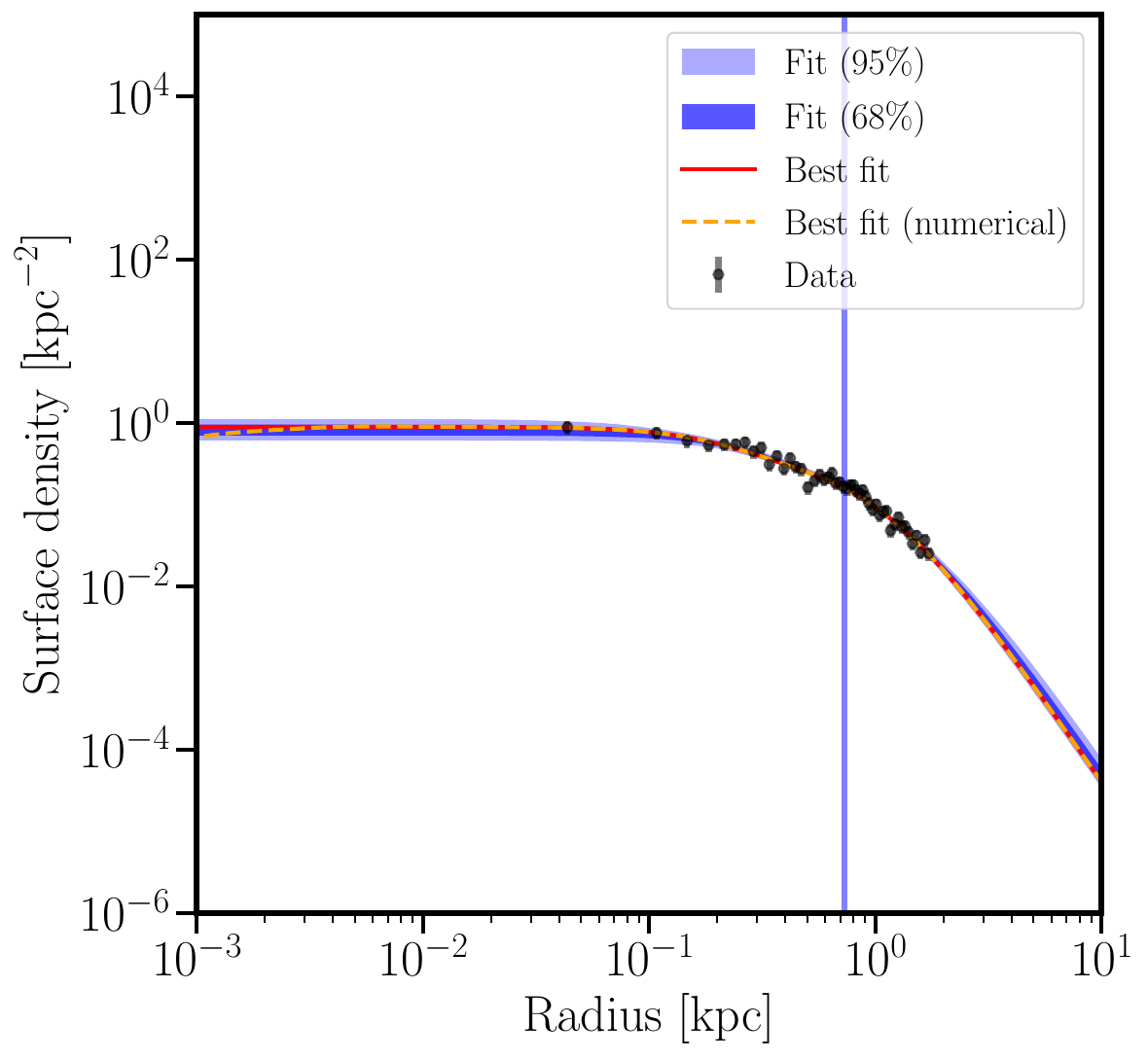}
    \caption{Surface brightness profile of Andromeda XXIII, as fit by \texttt{Binulator}. The black circles are binned data, with uncertainties being represented by the size of each marker. The vertical blue line is the estimated half-light radius as determined by \texttt{Binulator}. An artificial cut of photometric data was performed at 1800 pc to prevent the fitting of background stars.}
    \label{fig:And23SBP}
\end{figure}

Figures \ref{fig:And6SBP} and \ref{fig:And23SBP} show the surface brightness profiles as fit by \texttt{Binulator}. Each figure contains binned data for the respective system, which have been fit by the program for an estimated half-light radius. These plots also provide visual validation of the size of the dwarf, specifically at which point it becomes indistinguishable from the background. As noted in $\S$ \ref{subsec:massmodels}, there was an artificial cut performed on the photometric data to avoid erroneous fitting of background stars. This cut was performed at 900 pc for Andromeda VI and 1800 pc for Andromeda XXIII.

\begin{figure*}
    \centering
    \includegraphics[width = \textwidth]{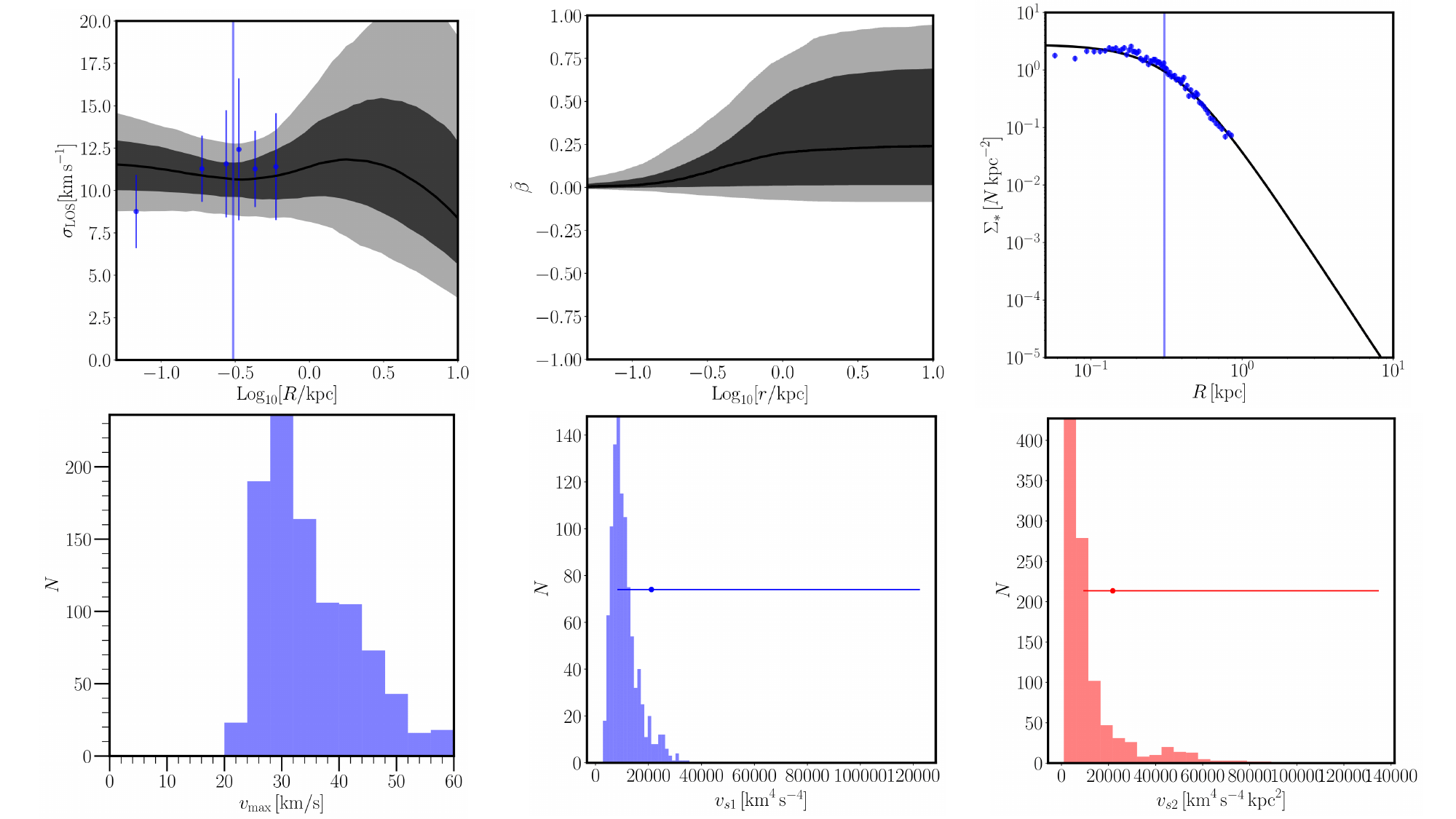}
    \caption{\gravsphere ~diagnostic plots for Andromeda VI. \textbf{Top Left}: Projected line-of-sight velocity dispersion ($\sigma_{\rm{LOS}}$). A 68\% confidence interval is denoted by the dark gray contour, while a 95\% confidence interval is denoted by the light gray contour. \textbf{Top Middle}: Symmetrized velocity anisotropy profile ($\tilde\beta$), with the confidence interval contours being the same at the line-of-sight plot. \textbf{Top Right}: Tracer surface density profile ($\Sigma_{\rm{*}}$). The blue points and bars are the data and their associated error, respectively. The vertical blue line in the line-of-sight and tracer surface density plots represent the calculated half-light radius ($r_{\rm{h}}$ by \gravsphere. \textbf{Bottom Left}: Maximum circular velocity ($v_{\rm{max}}$) estimation. \textbf{Bottom Middle}: Virial shape parameter 1 ($vs1$) estimation. \textbf{Bottom Right}: Virial shape parameter 2 ($vs2$) estimation. Both $vs1$ and $vs2$ are used in breaking the $\rho - \beta$ degeneracy when solving the spherical Jean's equation. The results of $vs1$ and $vs2$ are clearly poor for this analysis. It is explained further \citet{Read_2017} that tighter constraints on $\rho$ and $\beta$ are only achieved with proper motions, which are unavailable for And VI.}
    \label{fig:And6VelArray}
\end{figure*}

\begin{figure*}
    \centering
    \includegraphics[width = \textwidth]{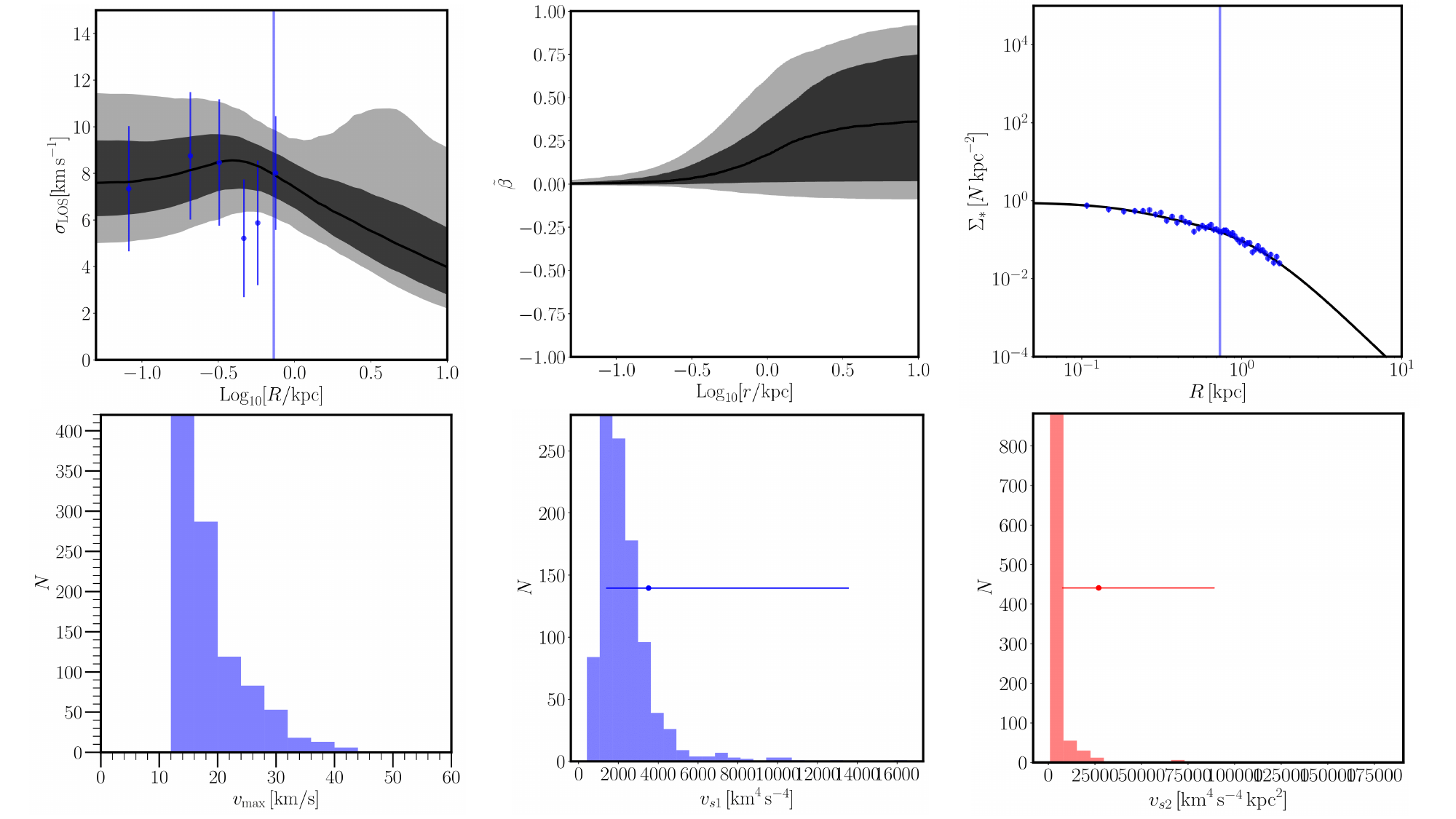}
    \caption{Same as Figure \ref{fig:And6VelArray} but for Andromeda XXIII. Proper motions are also unavailable for And XXIII, resulting in poor constraints on $vs1$ and $vs2$.}
    \label{fig:And23VelArray}
\end{figure*}

Figures \ref{fig:And6VelArray} and \ref{fig:And23VelArray} are \gravsphere ~diagnostic plots for Andromeda VI and Andromeda XXIII, respectively. These plots are used to understand how well the data are fit within the program and to visualize any errors estimated for each result. The plots included in this appendix are those for line-of sight velocity dispersion ($\sigma_{\rm{LOS}}$), symmetrized velocity anisotropy profile ($\tilde\beta$), tracer density profile ($\Sigma_{\rm{*}}$), maximum circular velocity ($v_{\rm{max}}$), and virial shape parameters 1 ($vs1$) and 2 ($vs2$).

\begin{figure}
    \centering
    \includegraphics[width=\columnwidth]{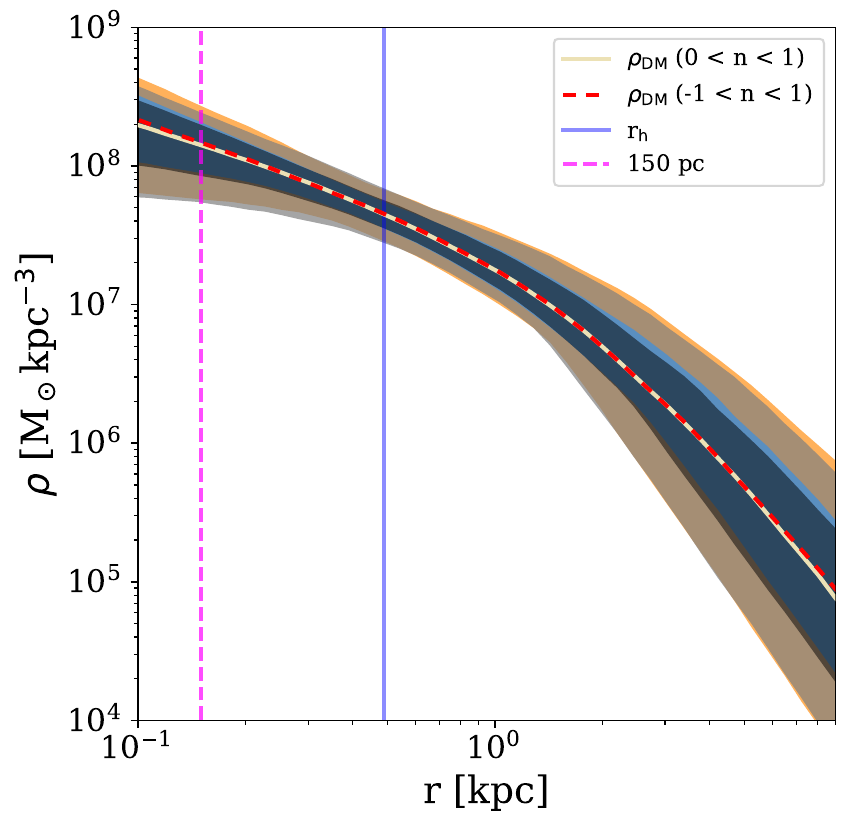}
        \caption{Comparison plot of estimated Andromeda VI dark matter density profiles when using differing prior ranges for the density shape parameter, $n$, in \gravsphere. The density profile from v1.5, which uses the range $0 < n < 1$, is shown as a solid cream line. This estimates a dark matter density at 150 pc (indicated by a vertical dashed pink line) of $\rho_{\rm{DM, v1.5}}$(150 pc) = (1.4 $\pm$ 0.5) $\times$ 10$^{8}$ M$_{\odot}$ kpc$^{-3}$. The $1\sigma$ and $2\sigma$ errors for the v1.5 profile are denoted by dark gray and light gray shaded regions, respectively. A profile estimated with the prior range $-1 < n < 1$, based on \gravsphere ~v1.0, is shown as a dashed red line. This presents slightly higher densities than the v1.5 profile at extreme radii. At 150 pc, the v1.0 density is estimated as $\rho_{\rm{DM, v1.0}}$(150 pc) = 1.5 $\substack{+0.6 \\ -0.5} \times 10^{8}$ M$_{\odot}$ kpc$^{-3}$. The $1\sigma$ and $2\sigma$ errors for the v1.0 profile are denoted by light blue and orange shaded regions, respectively. The main deviation from the errors estimated by v1.5 again appear at extreme radii, with little difference around 150 pc and no discernible difference at the half-light radius (shown as a vertical solid blue line). It can therefore be said that the differing prior range for $n$ between v1.0 and v1.5 is negligible in \gravsphere ~dark matter density estimations at 150 pc.}
    \label{fig:And6_prior_den_comp}
\end{figure}

Figure \ref{fig:And6_prior_den_comp} shows a comparison between the estimated dark matter density profiles of Andromeda VI depending on the shape parameter, $n$, prior range discussed in $\S$ \ref{sec:massmodelmethods}. All other priors were kept the same in an effort to isolate the effect of $n$ on the final density estimations. There exists a negligible difference between the profiles, with both estimated dark matter densities being virtually the same at 150 pc. \gravsphere ~v1.0 (plotted as a dashed red line) returns a value of $\rho_{\rm{DM, VI}}$(150 pc) = 1.5 $\substack{+0.6 \\ -0.5} \times 10^{8}$ M$_{\odot}$ kpc$^{-3}$, while v1.5 (plotted as a solid cream line) returns a value of $\rho_{\rm{DM, VI}}$(150 pc) = (1.4 $\pm$ 0.5) $\times$ 10$^{8}$ M$_{\odot}$ kpc$^{-3}$. Discrepancies only occur at extreme radii, with v1.0 presenting slightly higher densities and related errors than v1.5. In this, we find that there is no detriment in using one prior range over the other in our analysis of Andromeda VI. We would expect to find a similar result for Andromeda XXIII.

\newpage

\bsp	
\label{lastpage}
\end{document}